\documentclass{natureprintstyle}

\usepackage[final]{pdfpages}

\usepackage{psfig}

\usepackage{lscape}
\usepackage{amsmath}
\usepackage{amssymb}
\usepackage{color}
\usepackage{url}
\usepackage{ulem}
\usepackage{multirow}
\usepackage{graphics,graphicx}
\usepackage[affil-it]{authblk}
\usepackage{blindtext}

\usepackage{astjnlabbrev-nature} 

\usepackage{graphicx,amsmath}
\usepackage{caption}

\usepackage{astjnlabbrev-nature} 

\definecolor{asparagus}{rgb}{0.53, 0.66, 0.42}

\newcommand{\zpaper}{\ifmmode z{=}0.064 \else $z{=}0.064$\fi}
\newcommand{\mysobj}{AT\,2017bgt}
\newcommand{\Nmonths}{14}

\def \hbfittel {Palomar}
\def \hbfittelfull {Hale 200-inch telescope at the Palomar observatory}
\def \hbfitinst {DoubleSpec}
\def \hbfitdate {2017 March 4}
\def \hbfitfw {2,049}
\def \Loptfourthree {7.3}
\def \Mseven {1.8}

\newcommand{\kms}	{\ifmmode {\rm km\,s}^{-1} \else km\,s$^{-1}$\fi}
\newcommand{\cc}	{\ifmmode {\rm cm}^{-3}    \else cm$^{-3}$\fi}
\newcommand{\cmii}	{\ifmmode {\rm cm}^{-2}    \else cm$^{-2}$\fi}
\newcommand{\ergs}	{\ifmmode {\rm erg\,s}^{-1} \else erg s$^{-1}$\fi}
\newcommand{\ergcms}	{\ifmmode {\rm erg\,cm}^{-2}\,{\rm s}^{-1} \else erg\,cm$^{-2}$\,s$^{-1}$\fi}
\newcommand{\kev}	{\ifmmode {\rm keV} \else keV\fi}
\newcommand{\Msun}	{\ifmmode M_{\odot} \else $M_{\odot}$\fi}
\newcommand{\Msol}{\Msun}
\newcommand{\mpyr}{\ifmmode \Msun\,{\rm yr}^{-1} \else $\Msun\,{\rm yr}^{-1}$\fi}

\def\gsim{\;\rlap{\lower 2.5pt \hbox{$\sim$}}\raise 1.5pt\hbox{$>$}\;}
\def\lsim{\;\rlap{\lower 2.5pt \hbox{$\sim$}}\raise 1.5pt\hbox{$<$}\;}
\newcommand{  \Halpha   }{\ifmmode {\rm H}\alpha \else H$\alpha$\fi}
\newcommand{  \halpha   }{\Halpha}

\newcommand{  \Hbeta    }{\ifmmode {\rm H}\beta \else H$\beta$\fi}
\newcommand{  \hbeta    }{\Hbeta}
\newcommand{  \hb       }{\Hbeta}
\newcommand{  \hei      }{\ifmmode {\rm He}\,\textsc{i} \else He\,\textsc{i}\fi}
\newcommand{  \heii     }{\ifmmode {\rm He}\,\textsc{ii} \else He\,\textsc{ii}\fi}
\newcommand{  \HeIIuv   }{\ifmmode {\rm He}\,\textsc{ii}\,\lambda1640 \else He\,\textsc{ii}\,$\lambda1640$\fi}
\newcommand{  \HeIIop   }{\ifmmode {\rm He}\,\textsc{ii}\,\lambda4686 \else He\,\textsc{ii}\,$\lambda4686$\fi}
\newcommand{  \NIIopt   }{\ifmmode \left[{\rm N}\,\textsc{ii}\right]\,\lambda6584 \else [N\,\textsc{ii}]\,$\lambda6584$\fi}
\newcommand{  \nii      }{\ifmmode \left[{\rm N}\,\textsc{ii}\right]  \else [N\,\textsc{ii}]\fi}
\newcommand{  \niii     }{\ifmmode {\rm N}\,\textsc{iii} \else N\,\textsc{iii}\fi}
\newcommand{  \NIII     }{\ifmmode {\rm N}\,\textsc{iii}\,\lambda4640 \else N\,\textsc{iii}\,$\lambda4640$\fi}
\newcommand{  \niv      }{\ifmmode {\rm N}\,\textsc{iv}  \else N\,\textsc{iv}\fi}
\newcommand{  \NIVuv    }{\ifmmode {\rm N}\,\textsc{iv}\,\lambda1486 \else N\,\textsc{iv}\,$\lambda1486$\fi}
\newcommand{\OII}{\ifmmode \left[{\rm O}\,\textsc{ii}\right]\,\lambda3727 \else [O\,{\sc ii}]\,$\lambda3727$\fi}
\newcommand{\oiii}{\ifmmode \left[{\rm O}\,\textsc{iii}\right] \else [O\,{\sc iii}]\fi}
\newcommand{\OIII}{\ifmmode \left[{\rm O}\,\textsc{iii}\right]\,\lambda5007 \else [O\,{\sc iii}]\,$\lambda5007$\fi}
\newcommand{  \OIIIbf   }{\ifmmode {\rm O}\,\textsc{iii}\,\lambda3133 \else O\,\textsc{iii}\,$\lambda3133$\fi}
\newcommand{  \oiv      }{\ifmmode {\rm O}\,\textsc{iv}  \else O\,\textsc{iv}\fi}
\newcommand{  \OIVIR    }{\ifmmode {\rm O}\,\textsc{iv}\,25.9\,\mu {\rm m} \else O\,\textsc{iv}\,$25.9\,\mu$m\fi}
\newcommand{ \fwhm  }{\ifmmode {\rm FWHM} \else FWHM\fi} 
\newcommand{ \fwhb  }{\ifmmode {\rm FWHM}\left(\hb\right) \else FWHM(\hb)\fi}
\newcommand{  \Ledd     }{\ifmmode L_{\rm Edd} \else $L_{\rm Edd}$\fi}
\newcommand{  \Luv      }{\ifmmode L_{1450} \else $L_{1450}$\fi}
\newcommand{  \Lop      }{\ifmmode L_{5100} \else $L_{5100}$\fi}
\newcommand{\fbol}{\ifmmode f_{\rm bol} \else $f_{\rm bol}$\fi}
\newcommand{\fbolwv}{\ifmmode f_{\rm bol}\left(\lambda\right) \else $f_{\rm bol}\left(\lambda\right)$\fi}
\newcommand{\fbolopt}{\ifmmode f_{\rm bol}\left(5100{\rm \AA}\right) \else $f_{\rm bol}\left(5100{\rm \AA}\right)$\fi}
\newcommand{  \mbh      }{\ifmmode M_{\rm BH} \else $M_{\rm BH}$\fi}
\newcommand{  \lledd    }{\ifmmode L/L_{\rm Edd} \else $L/L_{\rm Edd}$\fi}
\newcommand{  \Lbol     }{\ifmmode L_{\rm bol} \else $L_{\rm bol}$\fi}
\newcommand{  \mbul     }{\ifmmode M_{\rm bulge} \else $M_{\rm bulge}$\fi} 
\newcommand{  \mstar    }{\ifmmode M_{*} \else $M_{*}$\fi} 
\newcommand{  \mgal     }{\ifmmode M_{*} \else $M_{*}$\fi} 
\newcommand{  \mhost    }{\ifmmode M_{\rm host} \else $M_{\rm host}$\fi}

\newcommand{  \swift     }  {{\it Swift}}
\newcommand{  \nustar     }  {{\it NuSTAR}}
\newcommand{  \nicer     }  {{\it NICER}}

\newcommand  {\RBLR}        {\hbox{$ {R_{\rm BLR}} $}}
\def\arcsec{\hbox{$^{\prime\prime}$}}
\def\deg{\hbox{$^\circ$}}

\title{A new class of flares from accreting supermassive black holes}

\author{Benny Trakhtenbrot$^{1,2}$, % v
Iair Arcavi$^{2,3,4}$, % v
Claudio Ricci$^{5,6,7}$, % v
Sandro Tacchella$^{8}$, % v
Daniel Stern$^{9}$, 
Hagai Netzer$^{2}$,
Peter G.\ Jonker$^{10,11}$,
Assaf Horesh$^{12}$,
Julian\ E.\ Mej\'ia-Restrepo$^{13}$,
Griffin Hosseinzadeh$^{3,4}$, % v
Valentina Hallefors$^{3,4}$,
D. Andrew Howell$^{3,4}$, % v
Curtis McCully$^{3,4}$, % v
Mislav Balokovi\'{c}$^{8}$, % v
Marianne Heida$^{14}$, % v
Nikita Kamraj$^{14}$, % v
George B.\ Lansbury$^{15}$, % v
{\L}ukasz Wyrzykowski$^{16}$,
Mariusz Gromadzki$^{16}$, % v
Aleksandra Hamanowicz$^{16}$, 
S.\ Bradley Cenko$^{17,18}$,
David J.\ Sand$^{19}$, % v
Eric Y.\ Hsiao$^{20}$, % v
Mark M.\ Phillips$^{21}$,
Tiara R.\ Diamond$^{17}$, % v
Erin Kara$^{17,18}$,
Keith C.\ Gendreau$^{17}$, 
Zaven Arzoumanian$^{17}$,
Ron Remillard$^{22}$ 
} % authors

\begin{document}

\maketitle

\begin{affiliations}
\item %Benny Trakhtenbrot
Department of Physics, ETH Zurich, Wolfgang-Pauli-Strasse 27, CH-8093 Zurich, Switzerland.
\item %BT, Iair Arcavi (1/3), Hagai Netzer
School of Physics and Astronomy, Tel Aviv University, Tel Aviv 69978, Israel.
\item %Iair Arcavi (2/3) and also:
% D. Andrew Howell, Curtis McCully, Griffin Hosseinzadeh, Valentina Hallefors?
Department of Physics, University of California, Santa Barbara, CA 93106-9530, USA.
\item %Iair Arcavi (3/3 and the other UCSB ppl, but NOT Valentina?
Las Cumbres Observatory, 6740 Cortona Drive, Suite 102, Goleta, CA 93117-5575, USA.
\item %Claudio Ricci 1/3
N\'ucleo de Astronom\'ia de la Facultad de Ingenier\'ia, Universidad Diego Portales, Av. Ej\'ercito Libertador 441, Santiago, Chile
\item %Claudio Ricci 2/3
Chinese Academy of Sciences South America Center for Astronomy and
China-Chile Joint Center for Astronomy, Camino El Observatorio 1515, Las Condes, Santiago, Chile. 
\item %Claudio Ricci 3/3
Kavli Institute for Astronomy and Astrophysics, Peking University, Beijing 100871, China.
\item %Sandro Tacchella
Harvard-Smithsonian Center for Astrophysics, 60 Garden St, Cambridge, MA 02138, USA.
\item %Dan Stern
Jet Propulsion Laboratory, California Institute of Technology, 4800 Oak Grove Drive, MS 169-224, Pasadena, CA 91109, USA.
\item %Peter Jonker 1/2
SRON Netherlands Institute for Space Research, Sorbonnelaan 2, 3584 CA Utrecht, The Netherlands.
\item %Peter Jonker 2/2
Department of Astrophysics / Institute for Mathematics, Astrophysics and Particle Physics, Radboud University, P.O. Box 9010, 6500GL Nijmegen, The Netherlands.
\item %Assaf Horesh
Racah Institute of Physics, Hebrew University, Jerusalem 91904, Israel.
\item %Julian Mejia-Restrepo
European Southern Observatory, Casilla 19001, Santiago 19, Chile
\item % Caltech NuSTAR people:
Cahill Center for Astronomy and Astrophysics, California Institute of Technology, Pasadena, CA 91125, USA.
\item % G. Lansbury
Institute of Astronomy, University of Cambridge, Madingley Road, Cambridge, CB3 0HA, UK.
\item % OGLE people:
Warsaw University Astronomical Observatory, Al. Ujazdowskie 4, 00-478 Warszawa, Poland.
\item % S. Bradley Cenko 1/2
Astrophysics Science Division, NASA Goddard Space Flight Center, Greenbelt, MD 20771, USA.
\item % S. Bradley Cenko 2/2
Joint Space-Science Institute, University of Maryland, College Park, MD 20742, USA.
\item % David Sand
Department of Astronomy and Steward Observatory, University of Arizona, 933 N. Cherry Avenue, Tucson, AZ 85721, USA.
\item % Eric Hsiao
Department of Physics, Florida State University, 77 Chieftan Way, Tallahassee, FL 32306, USA.
\item % Mark M. Phillips
Carnegie Observatories, Las Campanas Observatory, Casilla 601, La Serena, Chile.
\item
MIT Kavli Institute for Astrophysics and Space Research, 70 Vassar Street, Cambridge, MA 02139, USA.
\end{affiliations}

\smallskip
\begin{abstract}
Accreting supermassive black holes (SMBHs) can exhibit variable emission across the electromagnetic spectrum and over a broad range of time-scales. 
The variability of active galactic nuclei (AGN) in the ultra-violet (UV) and optical is usually at the few tens of percent level over time-scales of hours to weeks\cite{Caplar2017_PTF_QSOs}. 
Recently, rare, more dramatic changes to the emission from accreting SMBHs have been observed, including tidal disruption events (TDEs)\cite{Gezari2012_TDE,Arcavi2014_TDEs_He,Holoien2014_TDE_AS14ae,Holoien2016_TDE_AS14li}, ``changing look'' AGN\cite{LaMassa2015_changing,MacLeod2016_CLAGN,Ricci2016_IC751_CLAGN,Runnoe2016_changing}, and other extreme variability objects\cite{Lawrence2016_PS1_slow_hypervar,Graham2017_extreme_var}. The physics behind the ``re-ignition'', enhancement, and ``shut-down'' of accretion onto SMBHs is not entirely understood. 
Here we present a rapid increase in ultraviolet-optical emission in the centre of a nearby galaxy marking the onset of sudden increased accretion onto a SMBH. The optical spectrum of this flare, dubbed \mysobj, exhibits a mix of emission features. Some are typical of luminous, unobscured AGN, but others are likely driven by Bowen fluorescence - robustly linked here, for the first time, with high-velocity gas in the vicinity of the accreting SMBH. 
The spectral features and increased UV flux show little evolution over a period of at least \Nmonths\ months. 
This disfavours the tidal disruption of a star as their origin, and instead suggests a longer-term event of intensified accretion. Together with two other recently reported events with similar properties, we define a new class of SMBH-related flares. 
This has important implications for the classification of different types of enhanced accretion onto SMBHs.
\end{abstract}

%%%%%%%%%%%%%%%%%%%%%%%%%%%%%%%%%%%%%%%%%%%%%%%%%%%%%%%%%%%%%%%%%%%%%%%
\paragraph*{}
\label{sec:intro}
\vspace{-0.5cm}
% \bigskip

%%%%%%%%%%%%%%%%%%%%%%%%%%%%%%%%%%%%%%%%%%%%%%%%%%%%%%%%%%%%%%%%%%%%%%%
\paragraph*{}
\label{sec:obj_and_obs}
\vspace{-0.5cm}

\mysobj\ was discovered by the All Sky Automated Survey for Supernovae (ASAS-SN\cite{Shappee2014_ASASSN_N2617}) as ASASSN-17cv on 2017 February 21 in the early-type galaxy 2MASX J16110570+0234002, at \zpaper\ (ref.~\citen{Kiyota2017_AT17bgt_Atel}; see Methods \S\ref{SM_sec_obs_img}). % ATel 10113
The long-term ASAS-SN optical data show that the total emission from the galaxy brightened by a factor of $\sim$50\% over a period of about two months, with half of the rise occurring within three weeks (see Supplementary Fig.~1).
Follow-up {\it Swift} observations show that the UV emission increased by a factor of $\sim$75 compared to {\it GALEX} data from 2004,
reaching a luminosity of $\nu L_\nu ({\rm NUV}) \simeq 8.9{\times}10^{44}\,\ergs$,
and that the X-ray emission increased by a factor of ${\sim}2{-}3$ compared to {\it ROSAT} data from 1990 August (see `Detection and photometric monitoring' and 'Archival multi-wavelength data' in Methods for details on all new and archival data).
The archival X-ray luminosity, of $L(2{-}10\,\kev) \simeq 7{\times}10^{42}\,\ergs$, and the archival UV to X-ray luminosity ratio are consistent with what is commonly observed in AGN (that is, a UV-to-X-ray spectral slope of $\alpha_{\rm ox} {\approx} -1.2$; see Methods \S\ref{SM_sec_mw_data}). 
Archival detections in the radio (from 1998; obtained by the Very Large Array) and in the mid-infrared (from 2010; obtained with the {\it Wide-field Infrared Survey Explorer}) can be accounted for by star formation in the host galaxy. 
Thus, \mysobj\ experienced a dramatic increase in its UV emission, accompanied by a smaller increase in optical and X-ray emission, sometime between 2004 and 2017.

\begin{figure*}
\centering
\includegraphics[width=0.450\textwidth]{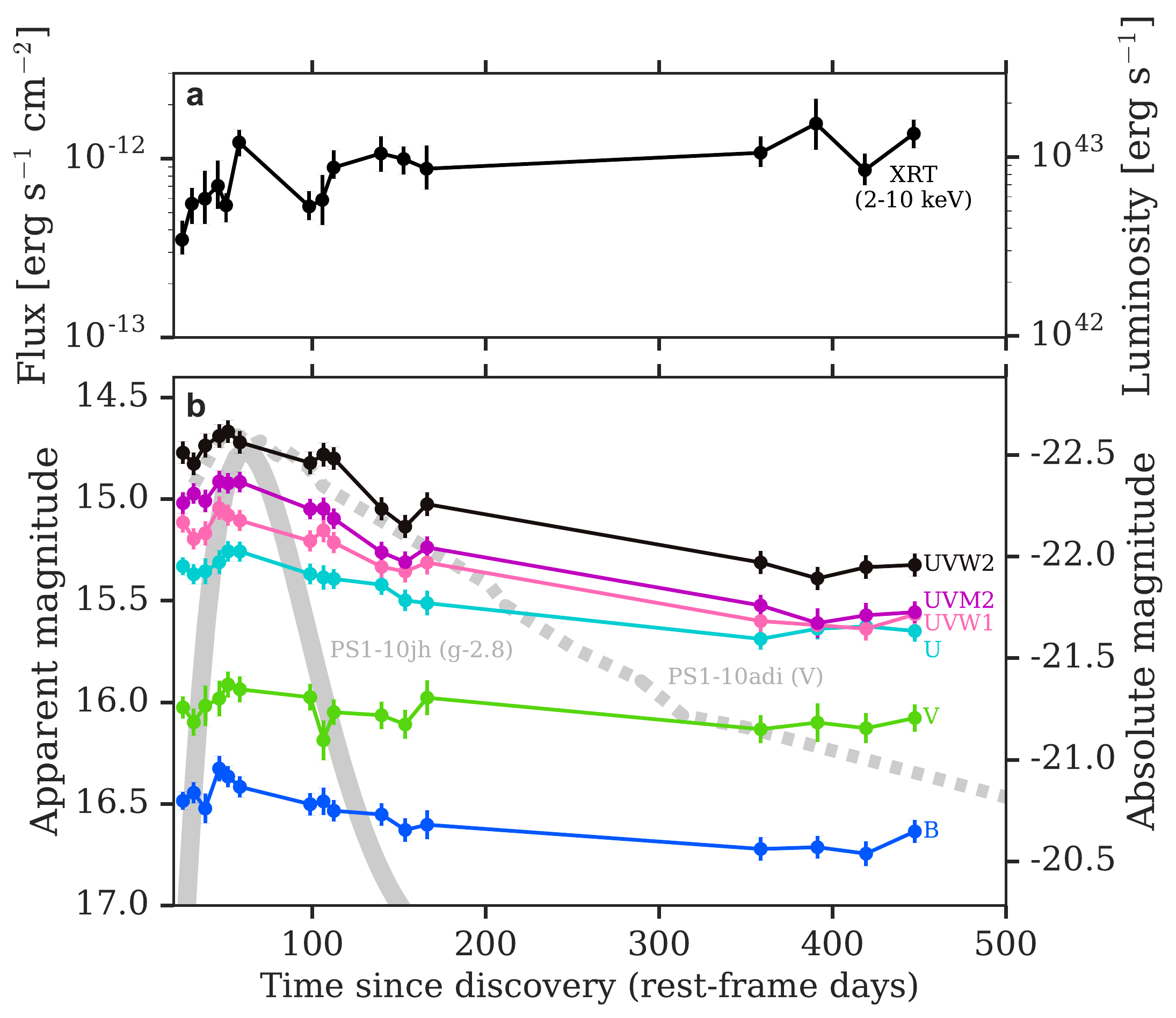}
\hfill
\includegraphics[clip, trim={0 0 0 0},width=0.525\textwidth]{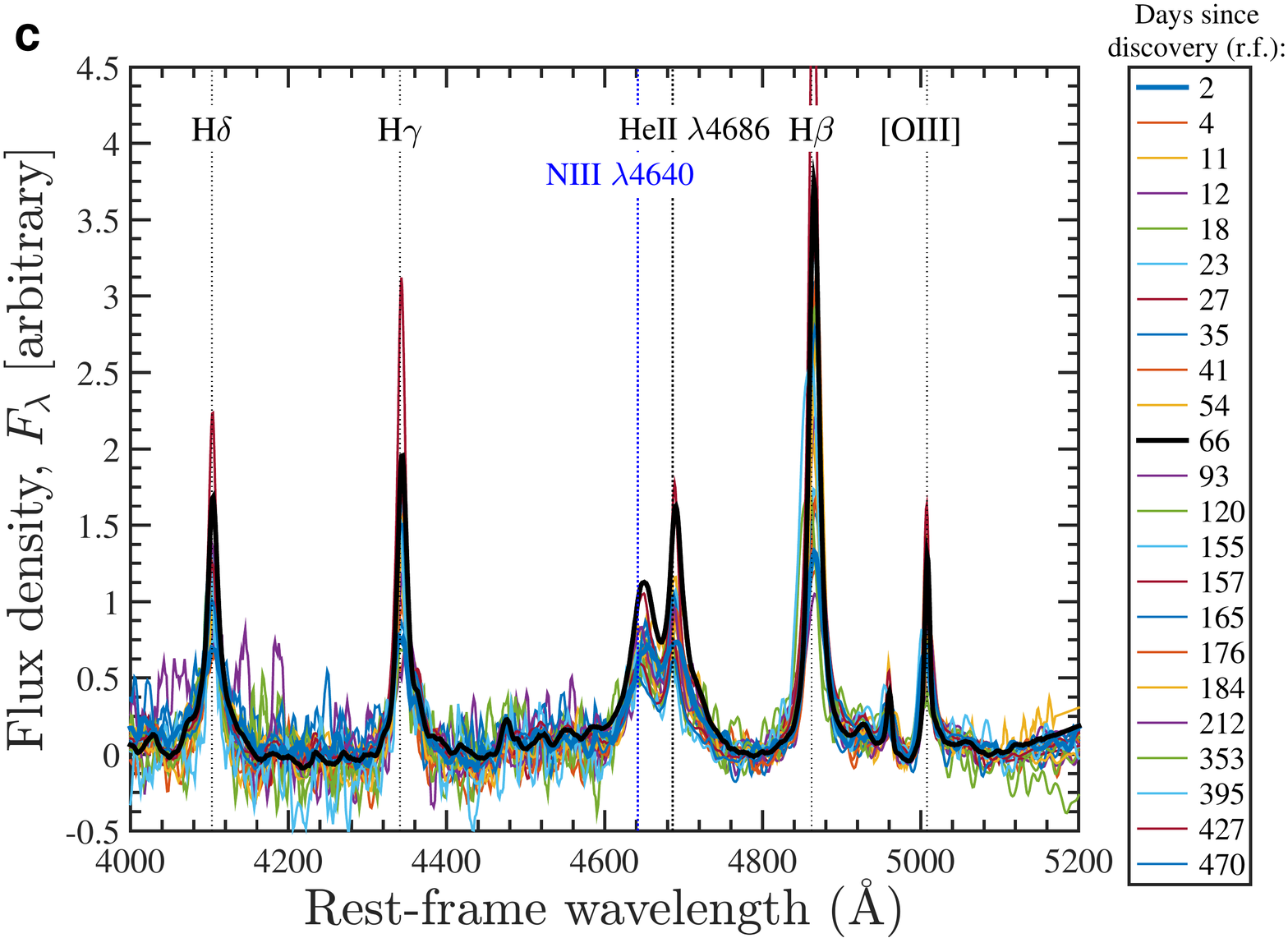}
\caption{\textbf{The persistence of X-ray and UV continuum, and of line emission in \mysobj.} 
Our monitoring of \mysobj, over the first \Nmonths\ months since the transient detection, suggests a roughly constant X-ray emission (panel {\bf a}). 
The post-detection UV-optical emission shows very little variability (panel {\bf b}; the different colours indicate magnitudes measured in different filters, in the AB system). 
The archival near-UV detection, not shown, is at an apparent magnitude of 19.6. 
This is markedly different than what is seen in TDEs or in other recently discovered classes of nuclear transients -- as exemplified by the optical light-curves of the TDE PS1-10jh (solid grey line, shifted arbitrarily and smoothed for clarity; ref.~\citen{Gezari2012_TDE}) and the slower nuclear event PS1-10adi (dashed grey line, plotted here to match its absolute $V$-band magnitudes; ref.~\citen{Kankare2017_PS10adi}).
Error bars in these light-curves are $1-\sigma$-equivalent.
Panel {\bf c} exhibits the persistent emission features in the optical spectra of \mysobj, including the broad, ${\approx}2,000\,\kms$-wide, Balmer lines and the double-peaked feature near 4680 \AA. 
The thick blue line highlights the first optical spectrum of the newly detected source, shown in Fig.~\ref{fig:opt_spec_comp_SDSS} (taken two days after discovery), while the thick black line highlights the high-resolution Keck/LRIS spectrum taken about two months after discovery.
All spectra have been smoothed by a 5-pixel boxcar filter, normalized at 5100 \AA, and continuum-subtracted. 
Time since detection is given in the side panel, in rest-frame (``r.f.'') days.
} % caption
\label{fig:img_spec_monitoring}
\end{figure*}

The X-ray spectral energy distribution of \mysobj, as determined from our new observations over the energy range of 0.3-24\,\kev, using \swift/XRT, \nustar, and \nicer, is broadly consistent with an unobscured AGN, dominated by a power-law with a photon index of $\Gamma{\simeq}1.9$ (see 'New X-ray data' in Methods and Supplementary Fig.~2). 
During the current UV-bright state, \mysobj\ exhibits UV-to-X-ray and UV-to-optical luminosity ratios that are much larger than what is typically seen in unobscured AGN. 
The ratio of monochromatic UV-to-X-ray luminosities, $L_{\nu}(2500\,{\rm \AA})/L_{\nu}(2\,\kev) \approx 6{\times}10^4$, is higher than the norm by a factor of at least $\sim$50 (i.e., $\alpha_{\rm ox}{\simeq}-1.9$; see ref.~\citen{Lusso2016_Lx_Luv}). 
The corresponding UV-to-optical ratio is $\gtrsim$3.5, which is higher than the norm by a factor of at least $\sim$5.5 (ref.~\citen{VandenBerk2001}).
Over the first \Nmonths\ months after discovery the UV and optical flux of \mysobj\ have shown very limited variability (Fig.~\ref{fig:img_spec_monitoring}), with a mild decline of $\lesssim$0.7 mag (factor of 2) in the UV, and $\lesssim$0.2 mag in the optical (without removal of an unknown amount of host contamination). 
The X-ray emission from \mysobj\ has been roughly constant during this period (Fig.~\ref{fig:img_spec_monitoring}a).

\begin{figure*}
\centering
% trim={<left> <lower> <right> <upper>}
\includegraphics[trim={0 0 8.7cm 0},clip,height=0.400\textwidth]{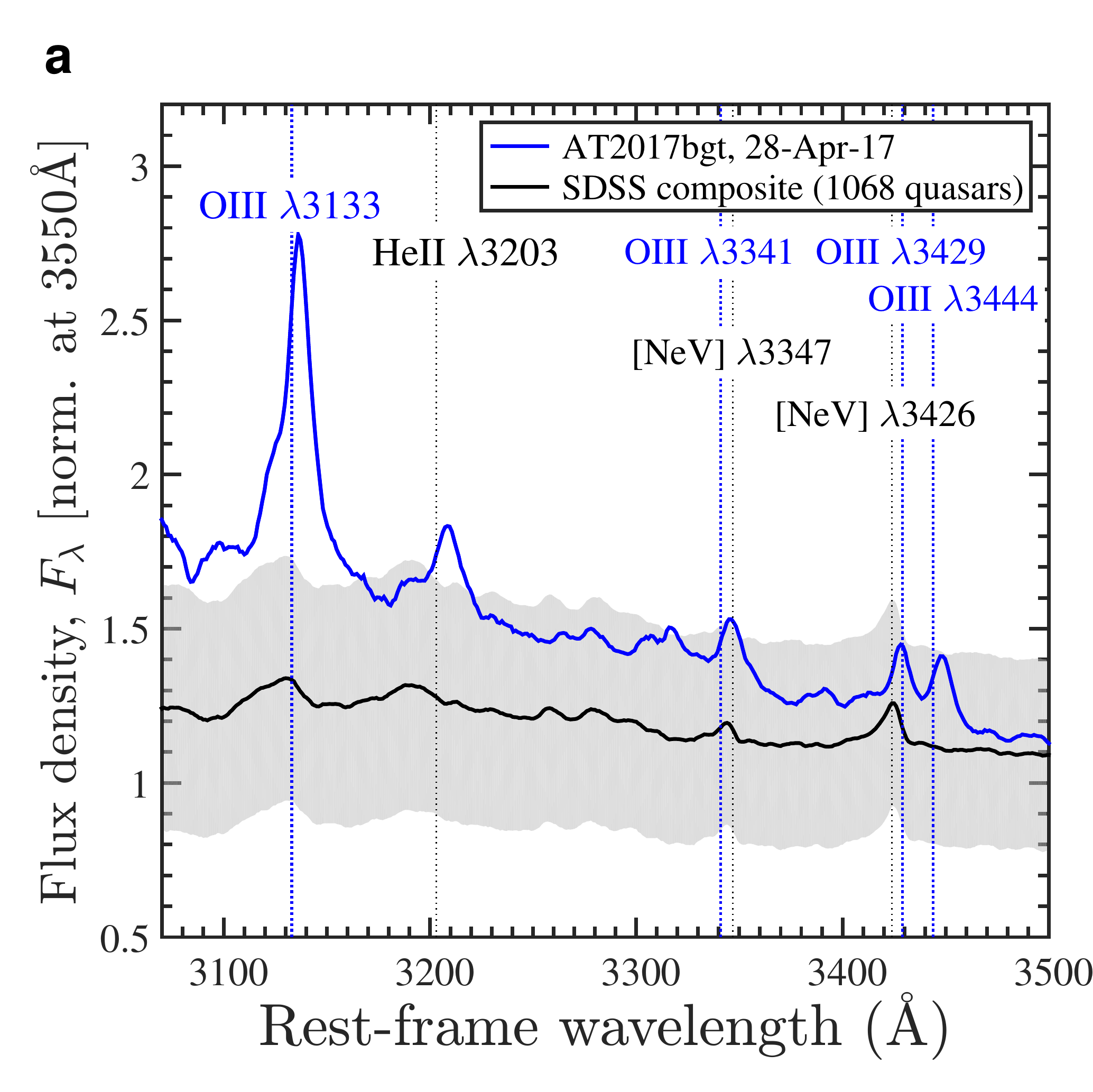}
\hfill
\includegraphics[height=0.400\textwidth]{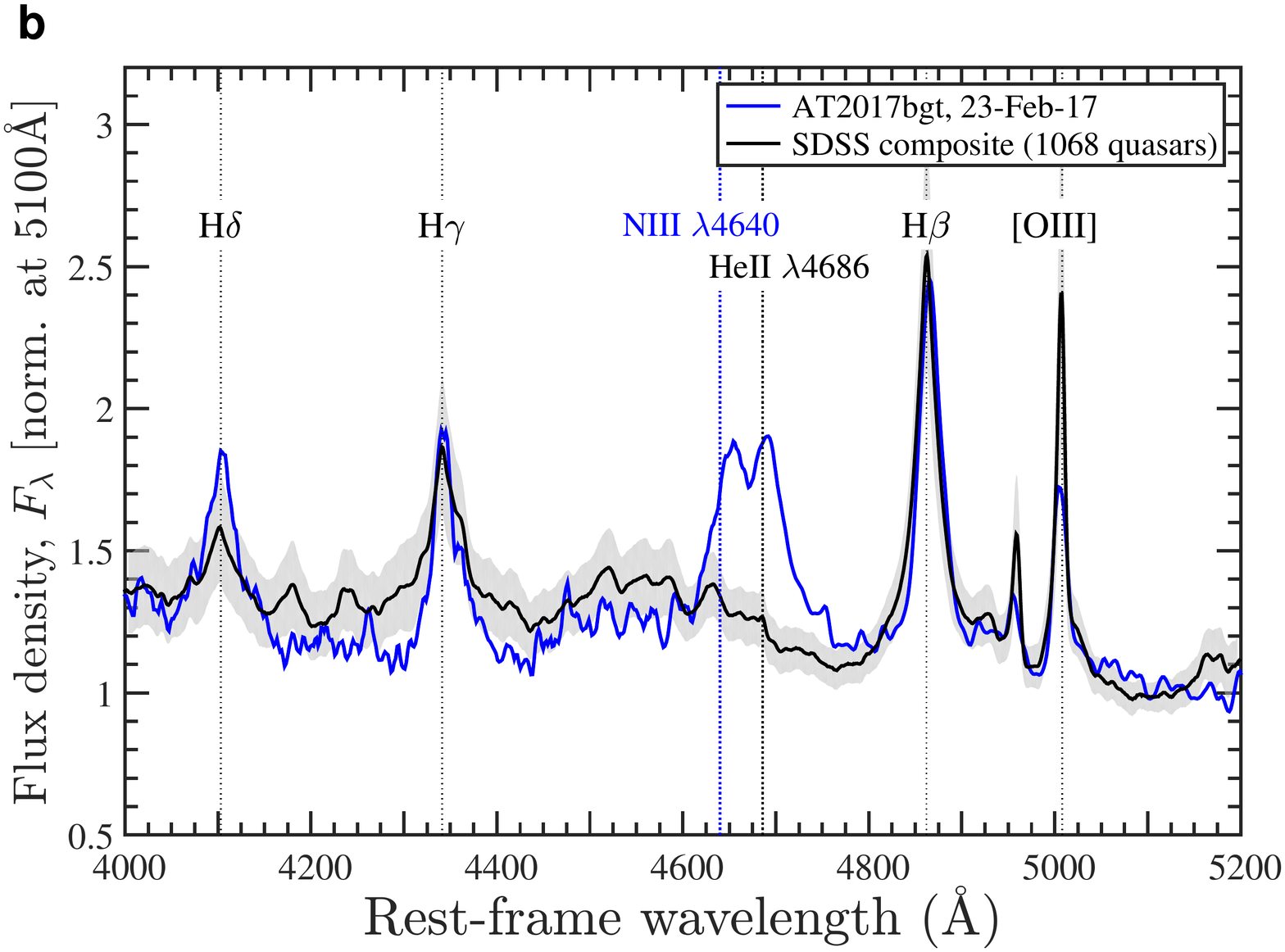}
\caption{
\textbf{The optical spectrum of \mysobj\ compared to known unobscured AGN.} %
Two different spectra of \mysobj\ (in blue), taken at different epochs within about two months from discovery (28 April and 23 February, 2017, for panels {\bf a} and {\bf b}, respectively),
are compared to a composite of 1068 broad-line AGN (quasars) with similar hydrogen emission line widths, taken from the Sloan Digital Sky Survey (see 'Optical spectroscopy' in Methods). 
The shaded region marks the median absolute deviation of the quasar sample.
The broad Balmer lines and narrow forbidden [O\,{\sc iii}]\,$\lambda\lambda4959,5007$ lines of \mysobj\ are similar to those seen in AGN.
On the other hand, the prominent
double-peaked emission feature near 4680\AA\ (panel {\bf b}), 
the prominent \OIIIbf\ and He\,{\sc ii} $\lambda3203$ lines, 
and the weaker O\,{\sc iii} $\lambda\lambda3341,3429,3444$ lines (panel {\bf a}), are not seen in such AGN.
These features indicate an atypically strong source of high-energy (UV) radiation that drives intense \heii\ emission, which in turn drives the O\,{\sc iii} and \niii\ emission lines through Bowen fluorescence (marked in blue). 
}
\label{fig:opt_spec_comp_SDSS}
\end{figure*}

We obtained repeated optical spectroscopy of \mysobj, starting two days after discovery.
The spectra display many features typical of unobscured AGN, but also some features that have never been clearly identified in such systems before 
(see 'Optical spectroscopy' in Methods and Fig.~\ref{fig:opt_spec_comp_SDSS}).
Among the features typical of AGN, the spectra exhibit prominent, symmetrical, single-peaked hydrogen Balmer emission lines, with full-width at half-maximum of $\fwhm {\approx} 2000\,\kms$. 
In unobscured, broad-line AGN (i.e., quasars), such emission lines are thought to originate from partially ionized gas with densities of order $10^{10}\,\cc$. 
\mysobj\ also exhibits weaker and narrower, ${\sim}500\,\kms$-wide, forbidden emission lines of [O\,{\sc iii}]\,$\lambda\lambda4959,5007$ and [N {\sc ii}]\,$\lambda\lambda6548,6584$, which are also common in AGN.
Indeed, these forbidden narrow lines are already present in the earliest optical spectra, obtained within days from the transient detection, and indicate 
the presence of an AGN-like ionizing source
(see Supplementary Figs.~3 and 4, and 'Optical spectroscopy' in Methods).
Given that the light-travel time to the corresponding (narrow) line-emitting region is of order $\gg$100 years\cite{Bennert2002_NLR_RL,Mor2009}, and the fact that the archival UV and X-ray data are consistent with AGN-like continuum emission, the system was most likely harbouring an actively accreting supermassive black hole (SMBH) well before the optical flux increase that triggered the transient detection.

Assuming that the optical and X-ray continuum emission from \mysobj\ indeed originates from the vicinity of a standard accretion flow onto a SMBH, as in AGN, and using standard AGN scaling relations, we infer a mass accretion rate of $\dot{M} {\sim} 0.04{-}0.11 \,\mpyr$; a broad line region (BLR) size of $\RBLR{\approx}20{-}30$ light-days (${\sim}2\times10^{-2}$ pc); and, combined with an \hbeta\ line width of $\fwhb{\simeq}2,050,\kms$, a SMBH mass of $\mbh {\approx} 1.8{\times}10^7\,\Msol$, which in turn results in an Eddington ratio of $\lledd {\sim} 0.08{-}0.21$ 
(see 'Determination of key SMBH properties' in Methods for details and Supplementary Fig.~5; key measured and derived properties are listed in Supplementary Table~1).
The intense UV luminosity, on the other hand, could imply $\lledd {\sim} 1.4$, and further suggests that the line-emitting region does not necessarily follow standard AGN scaling relations (see 'Determination of key SMBH properties' in Methods for details). 
Given the width of the Balmer lines and the high accretion rates suggested by the UV-based rough estimates of $\dot{M}$ and \lledd, 
the currently observed AGN-related properties of \mysobj\ are consistent with those of many narrow-line Seyfert 1 (NLSy1) galaxies.

Most importantly, the optical spectroscopy of \mysobj\ presents several strong emission features that are {\it not} seen in AGN (Fig.~\ref{fig:opt_spec_comp_SDSS}), namely the strong \OIIIbf\ emission line and the double-peaked emission feature near 4680\AA, all with widths consistent with that of the broad component of the \hbeta\ emission line.
The redder of the two peaks near 4680 \AA\ coincides with the \HeIIop\ transition, which in AGN typically exhibits line intensity ratios relative to \hbeta\ of   $F(\heii)/F(\hbeta) \leq 0.05$, but here is seen with $F(\heii)/F(\hbeta) \approx 0.5$ (see ref.~\citen{VandenBerk2001}).
The width and intensity of these spectral features remain roughly constant during our follow-up spectroscopic observations spanning over \Nmonths\ months from discovery (Fig.~\ref{fig:img_spec_monitoring}c and Supplementary Fig.~3).

The second feature, centred at 4651.6 \AA, cannot be associated with a second peak of \heii\ emission, originating from a disk-like configuration of the \heii-emitting gas, since the other optical and near-IR (NIR) emission lines, including several \heii\ transitions, are single-peaked (the NIR helium lines are also exceptionally strong compared to typical AGN; see 'NIR spectroscopy' in Methods and Supplementary Fig.~6).
This feature may instead be associated with the \NIII\ emission line.
Some recent studies have suggested that weak emission from this transition (sometimes noted as a ``Wolf-Rayet'' feature) may be present in the spectra of some TDEs\cite{Gezari2015_PS1_10jh,Brown2018_iPTF16fnl,Brown2017_AS14li_longterm}, although significant line blending in those cases makes the identification less secure there.
While the \NIII\ line is not seen in AGN spectra (Fig.~\ref{fig:opt_spec_comp_SDSS}), it can be significantly enhanced by Bowen fluorescence, which would also produce \OIIIbf\ and other O\,{\sc iii} lines, as seen in our spectra.
In Bowen fluorescence (BF), photons emitted from the Lyman$\alpha$-like transition of \heii, at 303.783 \AA, excite certain states of O\,{\sc iii} and \niii, due to the wavelength proximity between the corresponding energy transitions\cite{Bowen1928_BF,WeymannWilliams1969_BF}.
The excited O\,{\sc iii} and \niii\ states lead to a cascade of transitions, that may be observed as emission lines in the UV-optical regime.
This process, and particularly strong \OIIIbf, \NIII, and \HeIIop\ emission, is well established to occur in some planetary nebulae and X-ray binaries\cite{Schachter1989_BF_ScoX1,KastnerBhatia1996_BF_compilation}.
It is, however, generally not seen in AGN, and only a few Seyfert galaxies were reported to have narrow ($<$1,000 \kms-wide) \OIIIbf\ BF lines\cite{WilliamsWeymann1968_BF,Schachter1990_Bowen_NLR}.
Here we present a case where BF emission lines from the BLR are associated with a (steadily) accreting SMBH, and specifically robust identification of broad O\,\textsc{iii} BF lines, and of the \NIII\ line (and indeed the \niii\ BF cascade) in such a system, which were first predicted decades ago\cite{Netzer1985_HeII}.

Comparing the $F(\heii)/F(\hbeta)$ and $F(\niii)/F(\heii)$ line intensity ratios seen in \mysobj\ to those predictions\cite{Netzer1985_HeII} suggests that the line-emitting gas is dense, with hydrogen number density of $n_{\rm H}\gtrsim 10 ^{9.5}\,\cc$, and has a high abundance of metals -- and in particular of nitrogen (likely exceeding 4 times solar). 
These gas densities are consistent with what is expected for the BLR in AGN, and high metallicities may be expected in extremely high luminosity and/or high \lledd\ AGN\cite{Shemmer2004}.
Thus, \mysobj\ suggests that the key missing ingredient for broad BF lines in accreting SMBHs is extremely intense UV continuum emission. 
However, more detailed radiative transfer calculations are required to reproduce the line ratios seen in \mysobj, and to link them to the enhanced UV emission.

%%%%%%%%%%%%%%%%%%%%%%%%%%%%%%%%%%%%%%%%%%%%%%%%%%%%%%%%%%%%%%%%%%%%%%%%%%%%%%%%%%%%%%%%%%%%%%%%%%%%%%%%%%%%%%%%%%%%%%%%%%%%%%%%%%%%%%
% \section{Discussion \& Conclusion}
\paragraph*{}
\label{sec:discussion}
\vspace{-0.5cm}

\begin{figure*}
\centering
\includegraphics[width=0.875\textwidth]{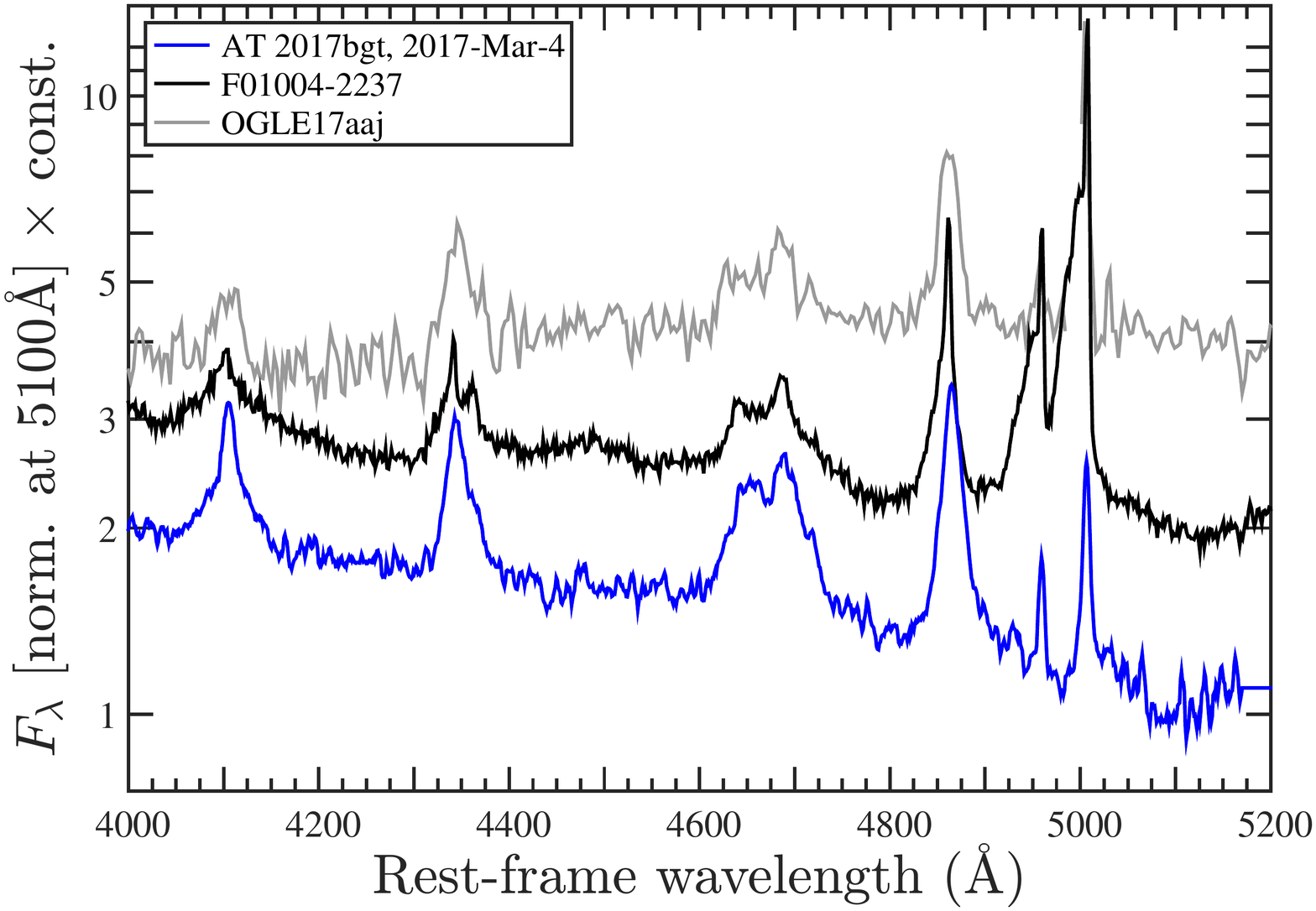}
\caption{\textbf{\mysobj\ as part of a new class of SMBH-related flares in galaxy nuclei.} 
The spectrum of \mysobj\ (in blue), taken 10 days after discovery is compared to the optical spectra of the transient in the nucleus of the galaxy F01004-2237 (black), recently claimed to be a tidal disruption event\cite{Tadhunter2017_TDE_ULIRG}; 
and the recently discovered transient OGLE17aaj (grey; refs.~\citen{Gromadzki2017_OGLEaaj_Atel,Wyrzykowski2017_OGLE_rep_Sept17}).
The spectra were normalized for presentation purposes.
All three spectra show emission lines that are characteristic of AGN, and a prominent double-peaked emission feature near 4680 \AA, not seen in AGN.
We interpret this double-peaked feature as a blend of broad \HeIIop\ and \NIII\ emission lines, driven by intense UV radiation and enhanced by Bowen fluorescence. 
}
\label{fig:opt_spec_comp_other}
\end{figure*}

\mysobj\ joins two recently reported transient events in galaxy centres that also exhibit a prominent, broad and double-peaked emission feature near 4680 \AA\ (Fig.~\ref{fig:opt_spec_comp_other}).
The first event was recently claimed to be a TDE in the active nucleus of the ultra-luminous infrared galaxy F01004-2237 (ref.~\citen{Tadhunter2017_TDE_ULIRG}). 
The classification of this event as a TDE mainly relied on the association of this spectral feature with \HeIIop, a line also seen in a class of optical and UV-bright TDEs\cite{Gezari2012_TDE,Arcavi2014_TDEs_He}, though the line profile in F01004-2237 is different from that of known TDEs (see below).
Another recently reported event, OGLE17aaj, shows hints of a persistent double-peaked emission feature on top of an otherwise normal optical AGN spectrum\cite{Wyrzykowski2014_OGLE_IV,Gromadzki2017_OGLEaaj_Atel,Wyrzykowski2017_OGLE_rep_Sept17} (M.G. et al., manuscript in preparation).
In both these cases, the dramatic increase in optical continuum emission was followed by a period of rather persistent continuum and line emission, over time-scales of several months -- similarly to \mysobj.
Moreover, the width of the Balmer emission lines in all three sources classifies them as NLSy1 AGN, which are commonly thought to be powered by highly accreting SMBHs (in terms of \lledd).
Two other recently reported nuclear transients showed slowly-evolving light curves: 
PS16dtm, which was claimed to be a TDE in a NLSy1 AGN\cite{Blanchard2017_PS16dtm};
and PS1-10adi, which was interpreted as a likely peculiar kind of supernova\cite{Kankare2017_PS10adi}. 
However, these two events showed no BF-related lines in their spectra.
Moreover, PS1-10adi (and similar events) show a significant flux decrease over periods of a few months, not seen in \mysobj. 
We therefore consider these two latter events to be unrelated to the new class identified here, that is: 
unobscured AGN-like spectra with extremely strong UV, \HeIIop, \NIII, and \OIIIbf\ emission, and long-term persistence of these flare features. 
We caution that, given the data in hand, we cannot rule out the possibility that these events may all be driven by rather similar SMBH fueling mechanisms.
Determining why some events show BF features and some do not requires more detailed modelling.

%The currently available data indicate
We propose that \mysobj, and likely similar events, are ``rejuvenated'' SMBHs that experienced a sudden increase in their UV-optical emission. 
This, in turn, enhanced the \heii\ Ly$\alpha$ line and initiated the BF cascades of \heii, \niii, and O\,{\sc iii}.
The extreme UV and BF emission differentiates such events from ``changing look'' AGN.
Given the likely light-travel timescales to the region emitting the broad Balmer lines (of order a few weeks), and the fact that they are seen in the earliest optical spectra we obtained, it is most likely that the BF emission features originate in a pre-existing BLR which was suddenly exposed to the intense ionizing UV emission whose spectral energy distribution is very different from those of normal AGN. 
It is therefore possible that, once the intense UV continuum emission would settle back to ``normal AGN'' levels, the BF features would disappear. 
This can be tested with on-going spectroscopic monitoring.
Alternatively, the BF features may be related to a newly launched outflow, perhaps driven by the sudden increase in accretion rate (see 'Relevant mechanisms for the long-lived UV flare' in Methods).
In either case, the enhanced metallicity suggested by the strong \NIII\ line may or may not be related to the fast increase in UV luminosity.

\begin{figure*}
\centering
\includegraphics[width=0.875\textwidth]{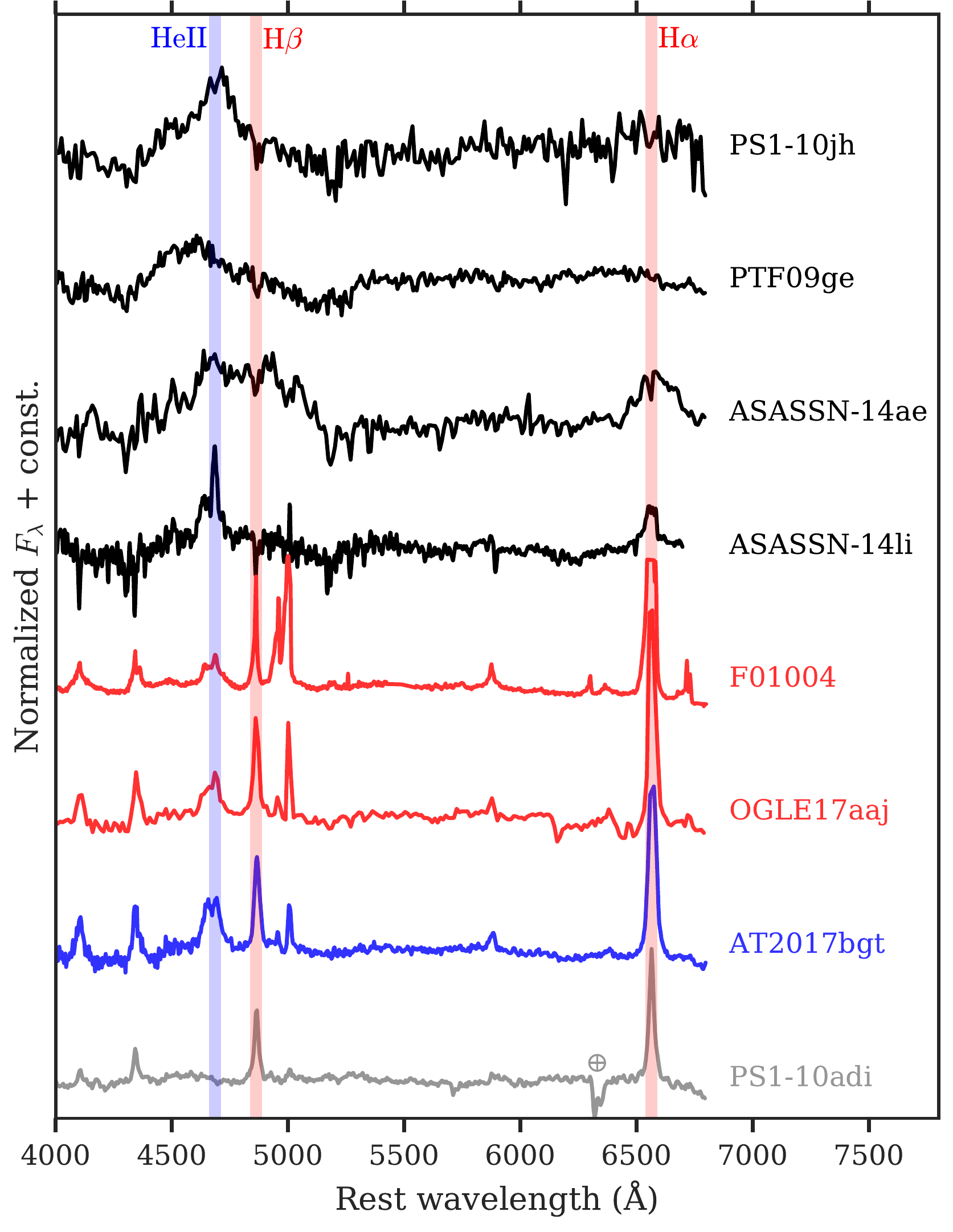}
\caption{\textbf{The broad emission features near \HeIIop\ in \mysobj, and similar objects, compared to other nuclear transients.}
Here we show the spectra of \mysobj\ (in blue) and the recently reported events in F01004-2237 (ref.~\citen{Tadhunter2017_TDE_ULIRG}) and OGLE17aaj (refs.~\citen{Gromadzki2017_OGLEaaj_Atel,Wyrzykowski2017_OGLE_rep_Sept17}; both in red), which we consider here to be part of a new class of nuclear transients. 
These are compared with spectra of four different TDEs (from refs.~\citen{Gezari2012_TDE,Arcavi2014_TDEs_He,Holoien2014_TDE_AS14ae,Holoien2016_TDE_AS14li}, in black), 
and of the luminous, slowly-evolving transient PS1-10adi (ref. \citen{Kankare2017_PS10adi}, in grey; the small cross marks a telluric feature).
All spectra are continuum-subtracted.
The broad emission feature near 4860 \AA\ in \mysobj, that we interpret as originating from \HeIIop\ and the Bowen fluorescence \NIII\ transitions, is significantly narrower than what is typically seen in most TDEs. 
Together with the very limited time evolution observed for these features in \mysobj, we disfavour a TDE interpretation for this event (and by association for the F01004-2237 and OGLE17aaj events as well).
}
\label{fig:heii_spec_comp_tdes}
\end{figure*}

The nature of the sudden UV and optical brightening event remains open.
The evidence for AGN-like activity in \mysobj\ (and the other events in this proposed class),
both before and after the UV-optical event, draws attention to processes related to (thin) accretion disks that feed SMBHs.
Given the very little time evolution of the continuum and line emission in \mysobj-like events, as well as their spectral properties, they are unlikely to be be driven by TDEs (at least not ones similar to those reported thus far; see Fig.~\ref{fig:heii_spec_comp_tdes}).
We discuss the relevance of the TDE interpretation, and mention several other possibly relevant mechanisms, in the Methods ('Relevant mechanisms for the long-lived UV flare').
Regardless of the nature of the dramatic increase in UV flux, the properties of \mysobj, and other similar transient events in galaxy nuclei, demonstrate that prominent, a broad Bowen fluorescence \NIII\ emission feature may be confused with the \HeIIop\ feature commonly seen in TDEs.
Such sources thus highlight the importance of long-term, multi-wavelength monitoring campaigns, in order to identify SMBH accretion that is driven purely by TDEs, and to distinguish it from other cases of SMBHs that experience a sudden enhancement, or indeed ``re-ignition'', of their accretion. 
We suspect that the discovery of \mysobj-like events might have been hampered so far, due to their possible association with previously-known AGN (see discussion in ref.~\citen{Kankare2017_PS10adi}).
Long term monitoring of \mysobj-like events, and indeed the discovery of additional events of this sort, could help illuminate their nature, as well as their role in SMBH growth.

%%%%%%%%%%%%%%%%%%%%%%%%%%%%%%%%%%%%%%%%%%%%%%%%%%%%%%%%%%%%%%%%%%
%% MAIN REFS
%%%%%%%%%%%%%%%%%%%%%%%%%%%%%%%%%%%%%%%%%%%%%%%%%%%%%%%%%%%%%%%%%%

% \bigskip
% \smallskip
\paragraph{References}
% \begin{thebibliography}{10}
%
% \vspace{-1.75cm}
% \bibliography{library}
% \bibliography{BIB_AT2017bgt_NatAst_20180528}

%%%%%%%%%%%%%%%%%%%%%%%%%%%%%%%%%%%%%%%%%%%%%%%%%%%%%%%%%%%%%%

%%%%%%%%%%%%%%%%%%%%%%%%%%%%%%%%%%%%%%%%%%%%%%%%%%%%%%%%%%%%%%%%%%
%% ACK, AUTHOR CONTRIBUTIONS, ETC
%%%%%%%%%%%%%%%%%%%%%%%%%%%%%%%%%%%%%%%%%%%%%%%%%%%%%%%%%%%%%%%%%%

\begin{addendum}

\item[Acknowledgements]

B.T.\ is a Zwicky Fellow.
I.A.\ is an Einstein fellow.
E.K.\ is a Hubble Fellow.
We thank N.\ Caplar, J.\ Guillochon, Z.\ Haiman, E.\ Lusso, and K.\ Schawinski for useful discussions.
We thank C.\ Tadhunter for providing the spectrum of the F01004-2237 transient and his helpful comments.
%
%
% ISSI:
Part of this work was inspired by discussions within International Team \#371, ``Using Tidal Disruption Events to Study Super-Massive Black Holes'', hosted at the International Space Science Institute in Bern, Switzerland.
We thank all the participants of the team meeting for their beneficial comments. 
%
%
% GRANTS:
Support for I.A.\ was provided by NASA through the Einstein Fellowship Program, grant PF6-170148.
C.R.\ acknowledges support from the CONICYT+PAI Convocatoria Nacional subvencion a instalacion en la academia convocatoria a\a~{n}o 2017 PAI77170080.
P.G.J.\ acknowledges support from European Research Council Consolidator Grant 647208.
A. Horesh acknowledges support by the I-Core Program of the Planning and Budgeting Committee and the Israel Science Foundation.
G.H., D.A.H., and C.M.\ acknowledge support from NSF grant AST-1313484.  
M.B.\ acknowledges support from the Black Hole Initiative at Harvard University, which is funded by a grant from the John Templeton Foundation.
G.B.L.\ acknowledges support from a Herchel Smith Research Fellowship of the University of Cambridge.
{\L}.W., M.G.\ and A.Hamanowicz acknowledge Polish National Science Centre grant OPUS no 2015/17/B/ST9/03167 to {\L}.W.
Research by D.J.S.\  is supported by NSF grants AST-1412504 and AST-1517649.
E.Y.H.\ acknowledges the support provided by the National Science Foundation under Grant No. AST-1613472 and by the Florida Space Grant Consortium.
%
%
% OBS:
%
This work makes use of observations from the Las Cumbres Observatory network.
This publication also makes use of data products from the Wide-field Infrared Survey Explorer.
{\it WISE} and {\it NEOWISE} are funded by the National Aeronautics and Space Administration.\\
This work made use of data from the \nustar\ mission, a project led by the California Institute of Technology, managed by the Jet Propulsion Laboratory, and funded by the National Aeronautics and Space Administration. 
We thank the \nustar\ Operations, Software and Calibration teams for support with the execution and analysis of these observations.
This research made use of the \nustar\ Data Analysis Software ({\it NuSTAR}DAS) jointly developed by the ASI Science Data Center (ASDC, Italy) and the California Institute of Technology (USA).\\
We thank the {\it Swift}, \nustar, and \nicer\ teams for scheduling and performing the target-of-opportunity observations presented here 
on short notice. 
The LRIS spectrum presented herein was obtained at the W. M. Keck Observatory, which is operated as a scientific partnership among the California Institute of Technology, the University of California, and the National Aeronautics and Space Administration. 
The Observatory was made possible by the generous financial support of the W.~M.\ Keck Foundation. 
We recognize and acknowledge the very significant cultural role and reverence that the summit of Mauna Kea has always had within the indigenous Hawaiian community. 
We are most fortunate to have the opportunity to conduct observations from this mountain.\\
These results made use of the Discovery Channel Telescope at Lowell Observatory. 
Lowell is a private, non-profit institution dedicated to astrophysical research and public appreciation of astronomy and operates the DCT in partnership with Boston University, the University of Maryland, the University of Toledo, Northern Arizona University and Yale University. 
The upgrade of the DeVeny optical spectrograph has been funded by a generous grant from John and Ginger Giovale.\\
The FLAMINGOS-2 spectrum was obtained at the Gemini Observatory under program GS-2017A-Q-33 (PI: Sand), which is operated by the Association of Universities for Research in Astronomy, Inc., under a cooperative agreement with the NSF on behalf of the Gemini partnership: the National Science Foundation (United States), the National Research Council (Canada), CONICYT (Chile), Ministerio de Ciencia, Tecnolog\'{i}a e Innovaci\'{o}n Productiva (Argentina), and Minist\'{e}rio da Ci\^{e}ncia, Tecnologia e Inova\c{c}\a~{a}o (Brazil).

%

%%%%%%%%%%%%%%%%%%%%%%%%%%%%%%%%%%%%%%%%%%%%%%%%%%%%%%%%%%%%%%
\item[Author contributions] 
 
B.T.\ and I.A.\ led the data collection, analysis and interpretation, as well as the manuscript preparation.
C.R.\ performed the analysis and modelling of archival and new X-ray data. 
S.T.\ performed the morphological and SED modelling of the host galaxy.
% to derive stellar population properties of the host galaxy. 
D.S., M.B., M.H., N.K., and G.B.L.\ took part in obtaining and calibrating the Palomar and Keck spectra.
H.N. contributed to the identification and interpretation of the Bowen fluorescence spectral features. 
P.G.J. and A. Horesh contributed to the interpretation of multi-wavelength data and to pursuing follow-up observations.
J.E.M-R. contributed to the analysis of optical spectra.
G.H., V.H., and C.M. contributed to collecting, calibrating and analysing the Las Cumbres Observatory and {\it Swift}/UVOT data.
D.A.H.\ helped schedule and monitor the data from the Las Cumbres Observatory.
{\L}.W., M.G., and A.\ Hamanowicz contributed to NIR line identification and provided the optical spectrum of OGLE17aaj.
S.B.C.\ provided the the DCT spectrum.
D.S.\ provided the the Gemini-South/FLAMINGOS-2 NIR spectrum.
E.Y.H., M.M.P., and T.D.D.\ provided the Magellan/FIRE NIR spectrum.
E.K. contributed to the X-ray data analysis and interpretation.
K.C.G., Z.A., and R.R. contributed to the \nicer\ data acquisition and calibration.

 \item[Additional information]
 {\bf Supplementary information} is available for this paper.\\ 
 {\bf Correspondence and requests for materials} should be addressed to B.T. (email: benny@astro.tau.ac.il).
 
 \item[Competing Interests]
 The authors declare that they have no competing financial interests.
\end{addendum}
%%%%%%%%%%%%%%%%%%%%%%%%%%%%%%%%%%%%%%%%%%%%%%%%%%%%%%%%%%%%%%

\clearpage

%%%%%%%%%%%%%%%%%%%%%%%%%%%%%%%%%%%%%%%%%%%%%%%%%%%%%%%%%%%%%%%%%%
%% FIGURES
%%%%%%%%%%%%%%%%%%%%%%%%%%%%%%%%%%%%%%%%%%%%%%%%%%%%%%%%%%%%%%%%%%

%%%%%%%%%%%%%%%%%%%%%%%%%%%%%%%%%%%%%%%%%%%%%%%%%%%%%%%%%%%%%%%%%%%%%%%
%%%  FIG: SWIFT LC and SPEC EVO %%%%%%%%%%%%%%%%%%%%%%%%%%%%%%%%%%%%%%
% \newpage
%%%%%%%%%%%%%%%%%%%%%%%%%%%%%%%%%%%%%%%%%%%%%%%%%%%%%%%%%%%%%%%%%%%%%%%

%%%%%%%%%%%%%%%%%%%%%%%%%%%%%%%%%%%%%%%%%%%%%%%%%%%%%%%%%%%%%%%%%%%%%%%
%%%  FIG: OPT SPEC COMP. TO SDSS %%%%%%%%%%%%%%%%%%%%%%%%%%%%%%%%%%%%%%
\clearpage
\newpage
%%%%%%%%%%%%%%%%%%%%%%%%%%%%%%%%%%%%%%%%%%%%%%%%%%%%%%%%%%%%%%%%%%%%%%%

%%%%%%%%%%%%%%%%%%%%%%%%%%%%%%%%%%%%%%%%%%%%%%%%%%%%%%%%%%
%%%  FIG: OPT SPEC COMP. OTHER OBJS %%%%%%%%%%%%%%%%%%%%%%%%%%%%%%%%%%%
\clearpage
\newpage
%%%%%%%%%%%%%%%%%%%%%%%%%%%%%%%%%%%%%%%%%%%%%%%%%%%%%%%%%%%%%

%%%%%%%%%%%%%%%%%%%%%%%%%%%%%%%%%%%%%%
%%%  FIG: OPT SPEC COMP. TO TDEs
%%%%%%%%%%%%%%%%%%%%%%%%%%%%%%%%%%%%%%
\clearpage
\newpage
%%%%%%%%%%%%%%%%%%%%%%%%%%%%%%%%%%%%%%%%%%%%%%%%%%%%%%%%%%%%%%%%%%%%%%%
\clearpage

%%%%%%%%%%%%%%%%%%%%%%%%%%%%%%%%%%%%%%%%%%%%%%%%%%%%%%%%%%%%%%%%%%%%%%%%%%%%%%%%%%%%%%%%%%%%%%%%%%%%%%%%%%%%%%%%%%%%%%%%%%%%%%%%%%%%%%

%%%%%%%%%%%%%%%%%%%%%%%%%%%%%%%%%%%%%%%%%%%%%%%%%%%%%%%%%%%%%%%%%%
% METHODS
%%%%%%%%%%%%%%%%%%%%%%%%%%%%%%%%%%%%%%%%%%%%%%%%%%%%%%%%%%%%%%%%%%

\clearpage
\newpage
\paragraph{Methods}
\label{sec:methods}

%%%%%%%%%%%%%%%%%%%%%%%%%%%%%%%%%%%%%%%
% \clearpage
% \newpage
\bigskip

% \makeatletter 
% \renewcommand{\thefigure}{S\@arabic\c@figure}
% \makeatother

% \makeatletter 
% \renewcommand{\thetable}{S\@arabic\c@table}
% \makeatother

\makeatletter 
\renewcommand{\thesection}{S\@arabic\c@section}
\makeatother

\makeatletter 
\renewcommand{\figurename}{Supplementary Figure}
\makeatother

\setcounter{section}{0}
\setcounter{figure}{0}
\setcounter{table}{0}

%\begin{center}
% \noindent \textbf{\huge Supplementary Information}
%\end{center}

%%%%%%%%%%%%%%%%%%%%%%%%%%%%%%%%%%%%%%%%%%%%%%%%%%%%%%%%%%%%%%%%%%%%%%%%%%%%%%%%%%%%%%%%%%%%%%%%%%%%%%%%%%%%%%%%%%%%%%%%%%%%%%%%%%%%%%
% \section{Observations and Data Analysis}
% \label{SM_sec_obs}

%%%%%%%%%%%%%%%%%%%%%%%%%%%%%%%%%%%%%%%%%%%%%%%%%%%%%%%%%%%%%%%%
\section*{Detection and Photometric Monitoring Observations}
\label{SM_sec_obs_img}

\noindent
The source \mysobj\ (J2000.0 coordinates $\alpha=$ 16:11:05.696, $\delta=$ +02:34:00.52) was first detected by the ASAS-SN survey, using the ASASSN-Brutus telescope, on 2017 February 21, 15:07:12 (${\rm JD}=2457806.13$; see refs.~\citen{Kiyota2017_AT17bgt_Atel,Stanek2017_AT17bgt_ASASSN}).
The nuclear source was discovered at a host-galaxy-subtracted $V$-band Vega magnitude of $17.2 \pm 0.1$. 
The previous 5-$\sigma$ non-detection from February 13 was at a magnitude of 17.5.\\
Supplementary Fig.~1 presents the long-term ASAS-SN $V$-band light-curve of the position of \mysobj, obtained using the ASAS-SN Light-Curve Server\cite{Shappee2014_ASASSN_N2617,Kochanek2017_ASASSN_LC}.
We note that the ASAS-SN $V$-band flux measurements are derived from apertures and calibration procedures that are significantly different from the \swift/UVOT measurements presented in Fig.~\ref{fig:img_spec_monitoring} (and described below). 
In particular, the ASAS-SN Light-Curve Server uses a large aperture, with a radius of 16\arcsec (compared to the 3.75\arcsec\ radius used for our new \swift/UVOT data; see below). 
As such, they mainly illustrate the transient nature of \mysobj, and can otherwise provide only combined flux measurements of the nuclear source and its host galaxy (with possible additional blending from nearby sources).\\

Evidently, a consistent trend of brightening began over one month prior to the detection of \mysobj. 
The automated ASAS-SN data suggest that the pre-event ASAS-SN $V$-band magnitude of the host galaxy was 15.85, from which we deduce that the nuclear event itself represented a brightening by a (peak) flux density corresponding to 16.7 magnitudes, or by $\nu L_\nu ({\rm opt.})\simeq 4.2\times10^{43}\,\ergs$. 
After reaching peak brightness, the transient began fading at a slow rate, dimming by roughly 0.3 mag at about 200 days after discovery (in the rest-frame).
Later photometry, spanning $\sim$300-400 days after discovery, reveals brightness levels that are consistent with the peak. 
This limited evidence for any significant post-flare flux variability is consistent with what is seen in the \swift/UVOT photometric monitoring observations (Fig.~\ref{fig:img_spec_monitoring}).

\smallskip
\noindent
Shortly after detection, we initiated a monitoring campaign of \mysobj\ using the UV Optical Telescope\cite{Roming2005_Swift_UVOT} on board the Neil Gehrels \swift\ observatory\cite{Gehrels2004_Swift}, which covers the wavelength range $\lambda\sim1600-6000$ \AA.
The most relevant band for comparison with the archival {\it GALEX} data (see below) is UVM2, with $\lambda_{\rm eff}=2228$ \AA.
\mysobj\ was initially observed over 12 epochs, separated by 5.5-43 days.
Four additional visits, separated by about a month, took place roughly a year after the transient detection (see Fig.~\ref{fig:img_spec_monitoring}b).
For each of these visits, we measured the flux in all bands through circular apertures with a diameter of 7.5\arcsec, to match the archival {\it GALEX} data (see below). 
We subtracted the sky flux using the same size aperture from an empty region near the transient. 
For host-galaxy subtraction, we assumed the archival {\it GALEX} flux measurement, discussed below.
A more up-to-date measurement of the host contribution will only be available once the transient fades completely. 
Intrinsic UV luminosities were calculated by taking into account a Milky Way extinction with a colour excess of $E(B-V)=0.069$ and assuming a Cardelli et al.\ extinction law (ref.~\citen{Cardelli1989}; see also ref.~\citen{Wyder2005_UVLF_GALEX}).
This resulted in a correction of 0.6 magnitudes (or a factor of 1.74).
We note that the Milky Way dust correction, and the host light subtraction were not applied to the UV-optical data shown in Fig.~\ref{fig:img_spec_monitoring}, which instead shows the raw measured fluxes of \mysobj.
Similarly to what is seen in the optical light-curve, the UV emission fades very slowly over the \Nmonths\ months of our \swift/UVOT monitoring, from about 14.9 to 15.6 AB magnitudes (i.e., about 50\% drop in flux).
After correcting for Milky Way extinction, this corresponds to a drop of $4.12\times10^{44}\,\ergs$ in luminosity, from $\nu L_\nu ({\rm NUV}){=}8.85{\times}10^{44}$ to $4.73{\times}10^{44}\,\ergs$.

The fading seen in both the UV and optical emission is much slower than what is typically seen in TDEs.
The only way in which the $\sim$0.7 magnitude decline in the UV light can be accounted for with a TDE light-curve of the form $L\propto(t-t_{\rm D})^{-5/3}$, is if the disruption time $t_{\rm D}$ was about two years before the discovery of \mysobj\ (observed frame, thus about 23 months in rest-frame), and is therefore inconsistent with the data.

%%%%%%%%%%%%%%%%%%%%%%%%%%%%%%%%%%%%%%%%%%%%%%%%%%%%%%%%%%%%%%%%
\smallskip
\section*{Archival multi-wavelength data}
\label{SM_sec_mw_data}

~~
\par

\noindent
The host galaxy of \mysobj\ is clearly detected in over 9 separate visits of the Panoramic Survey Telescope and Rapid Response System (Pan-STARRS), in all the 5 filters ($grizy$; with 9, 18, 26, 11, and 13 detections in each of these filters).
The stacked images, which have ``mean epochs'' ranging 2011 August 17 to 2013 June 8, and are publicly available through the Pan-STARRS database, reveal an early-type host galaxy, with no obvious sign of a central point source.

A full \texttt{GALFIT} analysis\cite{Peng2010_GALFIT_v3} of the stacked $r$-band image, which has a mean epoch of 2011 August 17 (i.e., about 5.5 years before the transient detection), suggests a bulge-dominated morphology.
For a single S\'{e}rsic fit, we find a S\'{e}rsic index of $n=5.0^{+0.9}_{-0.6}$. 
The uncertainties are estimated by varying the centre of the light profile within a box of 2$\times$2 pixels, following the approach described in ref.~\citen{Tacchella2015_SINS_ApJ}. 
Performing a bulge-disk decomposition with a fixed disk S\'{e}rsic index of $n_{\rm d}=1.0$ (pure exponential profile) gives a bulge-to-total ratio of $B/T=0.65^{+0.05}_{-0.06}$ with a bulge S\'{e}rsic index of $n_{\rm b}=3.7^{+0.8}_{-0.5}$. 
We further used \texttt{GALFIT} to assess the possible presence of a central point source.
Integrating over the residuals of the disk+bulge fit provides an $r$-band magnitude of $r_{\rm AB}=20.4^{+0.6}_{-0.5}$. 
Alternatively, adding a central point source component to the \texttt{GALFIT} model results in $r_{\rm AB}=19.75^{+0.3}_{-0.4}$, again much fainter than the bulge and disk components, which now have $r_{\rm AB}=15.8$ and $16.9$, respectively (and $B/T=0.73^{+0.06}_{-0.08}$). 
We conclude that the publicly available archival Pan-STARRS data shows no indication of a central point source.

\mysobj\ is associated with the {\it ROSAT} source 1RXS J161105.2+023350. 
The archival X-ray detection was obtained as part of the {\it ROSAT} All Sky Survey (RASS)\cite{Voges2000_RASS}, 
on 1990 August 9, 
when the source had a count rate of $(4.1\pm1.2)\times 10^{-2}\, {\rm ct\,s}^{-1}$. 
The {\it ROSAT} count rate can be converted into an observed $0.1-2.4$ \kev\ flux assuming that the photon index of the X-ray emission is $\Gamma=2$ within this band. 
This implies that the observed flux is $(5.4 \pm 1.5) \times 10^{-13}\,\ergcms$. 
The hydrogen line-of-sight column density of the Milky Way in the direction of the source is $5.34\times 10^{20}\,\cmii$, which provides an intrinsic $0.1-2.4\,\kev$ flux of $F(0.1{-}2.4\,\kev) {=} (1.1\pm 0.3) \times 10^{-12}\,\ergcms$.
The corresponding archival intrinsic luminosity in the same band is $L(0.1{-}2.4\,\kev){\approx} 10^{43}\,\ergs$, and in the 2--10\,keV range is $L(2{-}10\,\kev){\simeq} 5.3{\times}10^{42}\,\ergs$
(all luminosities and related quantities were calculated assuming a cosmological model with dark energy and matter densities of $\Omega_{\Lambda}=0.7$ and $\Omega_{\rm M}=0.3$, and a Hubble constant of $H_{0}=70\,\kms\,{\rm Mpc}^{-1}$). 
Considering the relation between $0.5-2\,\kev$ emission and star formation\cite{Ranalli2003_Lx_SFR}, a star formation rate of $\simeq 840\,\mpyr$ would be required to fully account for the X-ray emission in \mysobj. 
This extremely high value, which is two orders of magnitude larger than what we deduce from the archival data for the host galaxy of \mysobj\ (see below), means that {\it ROSAT} detected an AGN in this object.

\smallskip
\noindent
The position of \mysobj\ was observed in the 1.4 GHz band, as part of the FIRST radio survey, on 1998 July 17. The reported peak and integrated flux densities are $S_{\rm p} = 1.22\pm0.15$ mJy, and $S_{\rm int} = 0.99\pm0.15$ mJy (ref.~\citen{Helfand2015_FIRST_final_cat}).
This translates to a monochromatic radio luminosity of $\nu L_\nu({\rm 1.4\,GHz}) {\simeq} 1.7 \times 10^{38}\,\ergs$ 
Such a low radio luminosity can originate 
% If this radio emission originates 
purely from SF, with a corresponding star formation rate of about ${\rm SFR}{\simeq}7\,\mpyr$ (ref.~\citen{Yun2001_radio_IR_gals,Hopkins2003_SFR}), which is consistent with the UV-based SFR estimate derived from archival {\it GALEX} data (see below).
Indeed, the large majority of sources with such low radio luminosities are star-forming galaxies\cite{HeckmanBest2014_ARAA,Padovani2016_radio_rev}.
However, the aforementioned archival X-ray detection suggests that this may not be the case. 

\smallskip
\noindent
The area around \mysobj\ was also observed in the UV as part of the {\it GALEX} All-sky Imaging (AIS) survey, on 2004 May 17. The catalogued measured brightness of the corresponding source (GALEX J161105.7$+$023359) is 
$m_{\rm AB}({\rm NUV})=19.63 \pm 0.10$ in the NUV band ($\lambda_{\rm eff}=2274.4$\,\AA), and 
$m_{\rm AB}({\rm FUV})= 19.60 \pm 0.18$ in the FUV band ($\lambda_{\rm eff}=1542.3$\,\AA).
These are measured through a circular aperture with a diameter of 7.5\arcsec, which is preferable since it can be consistently applied to other imaging data for this source.
After correcting for Milky-Way extinction, the NUV brightness translates to a luminosity density of $L_\nu({\rm NUV}) = 8.8\times 10^{27}\,{\rm erg\,s}^{-1}\,{\rm Hz}^{-1}$.
Associating this luminosity with star formation in the host galaxy would translate to a star formation rate of ${\rm SFR}\simeq1.2\,\mpyr$ (for a Salpeter initial mass function; ref.~\citen{Kennicutt1998_SF_review}), which is within the range expected for a galaxy with the stellar mass of the host (see below; ref.~\citen{Salim2007_UV_SFRs}).
Alternatively, the UV continuum emission observed with {\it GALEX} can also be interpreted as originating from accretion onto the SMBH (i.e., an AGN).
Assuming no variability between the {\it ROSAT} and {\it GALEX} observations, we can calculate a UV-to-X-ray spectral slope, commonly defined through $F_\nu \propto \nu^{\alpha}$: 
\begin{equation}
\alpha_{\rm ox} \equiv \frac{\log(F_\nu[2\,\kev]/F_\nu[2500\,{\rm \AA}])}{\log(\nu[2\,\kev]/\nu[2500\,{\rm \AA}])} \approx -1.22 .
\label{SM_eq_aox}
\end{equation}
This value is in excellent agreement with what is expected from AGN with similar (UV) luminosities\cite{Just2007_Xray_hiL,Lusso2016_Lx_Luv}.

\smallskip
\noindent
Combining the {\it GALEX} data with the Pan-STARRS stacked optical images and 2MASS near-IR images, we obtain a UV-optical-near-IR spectral energy distribution (SED) of the host galaxy. All fluxes have been measured with the same aperture. The SED is fit assuming the stellar population synthesis templates of ref.~\citen{Bruzual2003}, and an exponentially declining star formation history ($\mathrm{SFR}\propto\exp(-t/\tau)$, where $\tau$ is a model parameter denoting a typical timescale).
This resulted in a stellar mass of $\mstar=2.7\times10^{10}\,\Msol$, a star formation rate of ${\rm SFR}=4.0\,\mpyr$, and a dust attenuation in the $V$-band of $\mathrm{A}_{\rm V}=0.5$.
Since the stellar mass does not depend on the UV-blue emission, we obtain an essentially identical mass estimate when refitting the SED but ignoring the {\it GALEX} data (i.e., if the archival UV detection is due to SMBH activity). 
However, the corresponding SFR drops to ${\rm SFR}\simeq0.02\,\mpyr$.

\smallskip
\noindent
We finally note that the area around \mysobj\ was observed with the {\it The Wide-field Infrared Survey Explorer (WISE)}\cite{Wright2010_WISE}, during 2010 February and August. 
The host galaxy is clearly detected in multiple scans, with $w(1-4)$ Vega magnitudes of (12.283,11.857,8.881,6.511), and no clear evidence for variability.
These suggest mid-IR colours that are consistent with star-forming galaxies\cite{Wright2010_WISE,Stern2012_MIR_AGN_WISE}.

%%%%%%%%%%%%%%%%%%%%%%%%%%%%%%%%%%%%%%%%%%%%%%%%%%%%%%%%%%%%%%%%
\bigskip
\section*{New X-ray data}
\label{SM_sec_Xrays}
~

\noindent
{\bf {\it Swift}/XRT observations and data.}
Our \swift\ monitoring campaign of \mysobj\ provided X-ray measurements with the X-Ray Telescope (XRT; ref.~\citen{Burrows2005_Swift_XRT}), covering the energy range $0.3-7\,\kev$.
The individual effective exposure times ranged between ${\sim}870{-}2400$ s.
The {\it Swift}/XRT data obtained in each of the 16 visits was analysed using the \textsc{xrtpipeline\,v0.13.4} within \textsc{heasoft\,v6.24} following standard guidelines, to provide a time-series of observed X-ray fluxes and SEDs.

The XRT SEDs were fitted with a simple power-law model. 
This resulted in photon indices in the range $\Gamma {\simeq} 1.9-2.5$ -- consistent with what is seen in low-redshift AGN\cite{Trakhtenbrot2017_BASS_GamX_LLedd}.
The X-ray flux in the 0.1--2.4\,keV band measured in these \swift/XRT observations, $F(0.1-2.4\,\kev) {=} (1.02{-}1.66) \times 10^{-12}\,\ergcms$, is ${\approx} 2-3$ times brighter than that recorded by {\it ROSAT} in the same energy range ($5.39{\times} 10^{-13}\,\ergcms$), back in 1990.
The new XRT measurements translate to (0.1--2.4) \kev\ luminosities of ${\approx} (1.0-1.6) {\times} 10^{43}\,\ergs$.
By assuming power-law continuum X-ray emission, we obtain 2--10\,\kev\ luminosities in the range $L(2{-}10\,\kev) {\approx} (3.5-15.5) {\times} 10^{42}\,\ergs$.
We adopt a nominal value of $L(2{-}10\,\kev)=1.2{\times}10^{43}\,\ergs$, corresponding to the 2017 April 24 \swift/XRT observation (see Supplementary Table~1).
Combining the series of \swift/XRT and UVOT measurements, we obtain UV-to-X-ray spectral slopes in the range $\alpha_{\rm ox}{\sim}(-2){-}(-1.7)$, with $\alpha_{\rm ox}=-1.8$ being both the value obtained for the April 24 \swift\ observation, and the median value.

We have also constructed a stacked X-ray SED from the first 12 epochs of \swift/XRT observations, corresponding to an effective exposure time of 20\,ks. 
While a model including a power-law and a blackbody component can adequately describe this stacked SED (resulting in a Cash statistic of $C-{\rm stat}=314.6$ for 318 degrees of freedom), it shows clear residuals around the Fe\,K$\alpha$ line region.
Adding a Gaussian component ('\textsc{zgauss}' in \texttt{XSPEC}) with the redshift fixed to systemic ($z{=}0.064$) improves the fit ($C-{\rm stat}=306.5$ for 316 d.o.f.), and results in a photon index of $\Gamma{=}2.20{\pm}0.17$ and a blackbody temperature of $k_{\rm B}T {=} 112^{+2}_{-3}$\,eV. 
The emission line has an energy consistent with the Fe\,K$\alpha$ fluorescent feature ($E{=}6.41^{+0.39}_{-0.15}$\,\kev), and is remarkably strong for an unobscured AGN, with an equivalent width of ${\rm EW}=900^{+494}_{-443}\,$eV. 
This is comparable with some of the strongest Fe\,K$\alpha$ lines known to date, for example, 
the NLSy1 galaxies IRAS\,13224$-$3809 (${\rm EW}\simeq2.4\,\kev$; ref.~\citen{Fabian2013_XMM_IRAS13224}) and 
1H\,0707$-$495 (${\rm EW}{\simeq} 0.97\,\kev$; ref.~\citen{Fabian2009_Fe_1H0707}), and
the obscured AGN IRAS\,00521$-$7054 (${\rm EW}\simeq 0.80\,\kev$; ref.~\citen{Ricci2014_Ka_IRAS00521}). 
We note, however, that the line strength significantly decreases in the higher-flux observations of \mysobj. 
Indeed, the emission line is not robustly detected in a stacked XRT SED that includes all epochs during which the flux was higher than the median, that is $F(2-10\rm\,keV){\gtrsim}10^{-12}\,\ergcms$. 
It is also not seen in the higher quality and much broader energy coverage X-ray spectrum we obtain using other facilities (see below).

%%%%%%%%%%%%%%%%%%%%%%%%%%%%%%%%%%%%%%%%%%%%%%%%%%%%%%%%%%%%%%%%

\noindent
{\bf \textit{NuSTAR} observations and data.}
\noindent
We observed \mysobj\ in the 3-24\,\kev\ energy range using the Nuclear Spectroscopic Telescope Array (\nustar; ref.~\citen{Harrison2013_NuSTAR}), on 2018 June 25 (starting UT16:11:09), for a total of $\sim$44 ks.
The {\it NuSTAR} data were processed using the {\it NuSTAR} Data Analysis Software \textsc{nustardas}\,v1.7.1 within \textsc{heasoft}\,v6.24, adopting the latest calibration files\cite{Madsen2015_NuSTAR_calib}. 
A circular region of 50\arcsec was used for source extraction, while the background was extracted from an annulus centred on the X-ray source, with an inner and outer radii of 60 and 120\arcsec, respectively.

%%%%%%%%%%%%%%%%%%%%%%%%%%%%%%%%%%%%%%%%%%%%%%%%%%%%%%%%%%%%%%%%
\noindent
{\bf \textit{NICER} observations and data.}
\noindent
The Neutron star Interior Composition ExploreR (\nicer; ref.~\citen{Gendreau2012_NICER}), mounted aboard the International Space Station, observed \mysobj\ on 2018 June 25 (ObsIDs: 1100440108, 1100440109; starting UT15:53:07 and 23:35:46, respectively) for the entirety of the \nustar\ observation, but was heavily affected by high background rates, leaving 18~ks of Good Time Intervals. 
The data were processed using the standard \nicer\ data analysis software ('DAS') v{\sc 2018-03-01\_V003} and were cleaned using standard calibration with '{\sc nicercal}' and standard screening with '{\sc nimaketime}'. 
To filter out high background regions, we made a cut on magnetic cut-off rigidity ({\sc COR\_SAX}~$>$ 2) and selected events that were detected outside the SAA. 
We selected events that were not flagged as ``overshoots'' or ``undershoots'' (EVENT FLAGS$=$bxxxx00). We also omitted forced triggers. We required pointing directions to be at least 30\deg\ above the Earth limb and 40\deg\ above the bright Earth limb. 
The cleaned events use standard ``trumpet'' filtering to eliminate additional known background events (using the tool {\sc nicermergeclean}). 
We estimated in-band background from the $13{-}15\,\kev$ and trumpet-rejected count-rates, and used this to select the appropriate background model from observations of a blank field. 
\nicer\ robustly detected \mysobj\ above the background level in the energy range $0.3{-}2\,\kev$.

%%%%%%%%%%%%%%%%%%%%%%%%%%%%%%%%%%%%%%%%%%%%%%%%%%%%%%%%%%%%%%%%
\noindent
{\bf Combined X-ray spectral analysis.}
\noindent
The combined, quasi-simultaneous $0.3-24\,\kev$ spectrum, including the \nustar and \nicer\ data of \mysobj, and a special short (400 s) \swift/XRT observation taken on 2018 June 26 (UT11:13:14), is shown in Supplementary Fig.~2. 
We fitted this spectrum with a spectral model consisting of a primary power-law and a blackbody component to account for the soft excess\cite{GierlinskiDone2004_soft_excess,Crummy2006_soft_excess}, as well as Galactic absorption (i.e., \textsc{tbabs(zpo+zbb)} in \texttt{XSPEC}).
This provides a good fit to the data, with $C$-Stat/$\chi^2{=}855$ for 918 degrees of freedom, and results in a blackbody temperature of $k_{\rm B} T {=} 145 {\pm} 4$ eV and a photon index of $\Gamma=1.94\pm0.07$. This slope is in very good agreement with that obtained by fitting the \nustar\ data alone ($\Gamma=1.95^{+0.09}_{-0.09}$). 
We also considered a cross-calibration constant ($C$) between the different instruments, which was fixed to 1 for \nustar\ Focal Plan Module A (FPMA), and was left free to vary for the other instruments. 
For \nustar/FPMB and \swift/XRT we obtained $C=1.07^{+0.09}_{-0.08}$ and $C=1.03^{+0.38}_{-0.30}$, respectively, while for \nicer\ we found a larger value ($C=1.48_{-0.18}^{+0.21}$), possibly due, at least in part, to variability, since the \nustar/\swift and \nicer\ observations were not completely overlapping.
Both of these measurements are consistent with what is found in local AGN\cite{Winter2012_X_SEDs_WAs,Ricci2017_BASS_Xray_cat}. 
We adopt the spectral parameters obtained from our fit to the combined, high-quality X-ray spectrum.
We note that this combined spectrum shows no clear sign of a Fe\,K$\alpha$ emission line, which is rather expected given that it was obtained when \mysobj\ was in a relatively high-flux state (see above).\\
%

%%%%%%%%%%%%%%%%%%%%%%%%%%%%%%%%%%%%%%%%%%%%%%%%%%%%%%%%%%%%%%%%
\section*{Optical spectroscopy.}
\label{SM_sec_obs_spec}

~~
\par

\noindent
{\bf Optical spectroscopic observations.}
Shortly after obtaining our classification spectrum, we initiated an intensive follow-up campaign using the Las Cumbres Observatory network of telescopes\cite{Brown2013_LCOGT}. 
% Brown et al. 2013 = http://adsabs.harvard.edu/abs/2013PASP..125.1031B
The spectra were obtained using the twin FLOYDS spectrographs mounted on the Las Cumbres Observatory 2-meter telescopes in the Haleakala, Hawaii,USA, and Siding Spring, Australia, observatories.    
These spectra cover a wavelength range of $\lambda{=}3,200{-}10,000$ \AA\ with a spectral resolution of $R {\equiv} \Delta\lambda/\lambda {=} 400{-}700$, and were obtained using 2\arcsec slits and 45-minute-long observations. 
The spectra were reduced using the PyRAF-based \texttt{floydsspec} pipeline (\url{https://github.com/svalenti/FLOYDS_pipeline/}).

Additional, higher-resolution and higher-S/N optical spectra were obtained on several occasions during the first six months of monitoring \mysobj\ through the spring and summer of 2017.
These include 
a March 3 spectrum taken with the DoubleSpec instrument on the Hale 200-inch telescope at the Palomar observatory (CA, USA);
a March 16 spectrum taken with the DeVeny instrument on the Discovery Channel Telescope, at the Lowell observatory (AZ, USA);
an April 28 spectrum taken with the LRIS instrument on the W. M. Keck telescope (HI, USA);
and another Palomar/DoubleSpec spectrum taken on July 28.
These spectra correspond to 10, 23, 66, and 157 days after the transient discovery, and are shown in Supplementary Fig.~3.
All these optical spectra were reduced following standard procedures. 
The higher-resolution spectra show the same features as seen in the ones obtained with the Las Cumbres Observatory, specifically the AGN-like Balmer, [O\,{\sc iii}], and [N\,{\sc ii}] emission lines, as well as the prominent, broad, and double-peaked emission feature near 4640 \AA.

To correct for the varying observing conditions during spectroscopic observations, we scaled all of our spectra to match the optical \swift/UVOT photometry, linearly interpolating the UVOT $V$-band data to the spectroscopic epochs and using standard synthetic photometry.
All our spectra are publicly available on the Weizmann Interactive Supernova data REPository (WISeREP\cite{YaronGalYam_2012_WISeREP}).

\smallskip
\noindent
{\bf Spectral analysis of the $\mathbf{\hbeta}$ complex.}
We decomposed the \hbeta\ spectral region of the \hbfitdate\ optical spectrum of \mysobj, taken with the \hbfitinst\ instrument mounted on the \hbfittelfull, following the methodology of ref.~\citen{TrakhtNetzer2012_Mg2}, with necessary adaptations. 
Specifically, we allowed for a much stronger broad \HeIIop\ emission line, and for an additional broad emission line to account for the blue peak seen in the optical spectra.
Each of these broad emission features is modelled by two Gaussian profiles, with widths forced to match those fit for the broad \hbeta\ line, and within the range $\fwhm= 2,000 - 10,000$ or $20,000\,\kms$.
The centres of these Gaussians are free to vary by $\pm1,500\,\kms$ relative to those of the broad \hbeta\ profile.
This procedure resulted in a satisfactory fit, shown in Supplementary Fig.~5, which indicates that the rest-frame equivalent widths of the broad \HeIIop\ and \NIII\ emission lines are ${\rm rEW}=27.7$ and 24.0 \AA, respectively.
We stress that the spectral decomposition procedure accounts for the much weaker, blended emission from ionized iron (see ref.~\citen{TrakhtNetzer2012_Mg2} for details).

\smallskip
\noindent
{\bf Composite spectrum construction.}
The SDSS composite shown in Fig.~\ref{fig:opt_spec_comp_SDSS} was constructed from 1068 quasars at $0.24 \leq z \leq 0.75$, drawn from the large sample analysed in ref.~\citen{TrakhtNetzer2012_Mg2}. 
The redshift range was chosen to assure that the spectral complexes surrounding both the \hbeta\ and \OIIIbf\ lines is within the SDSS coverage.
We further selected only those quasars for which the width of the broad component of the \hbeta\ line is in the range $1,800 {\leq} \fwhb {\leq} 2,200\,\kms$, comparable with what is measured in \mysobj.
The composite was then constructed by computing the geometrical mean in each 1 \AA\ (rest-frame) bin.

\smallskip
\noindent
{\bf Strong emission line ratio diagnostics.}
We used the four high-resolution spectra discussed above (Supplementary Fig.~3) to measure strong emission line ratio diagnostics, and specifically \OIII/\hbeta\ vs.\NIIopt/\halpha\ narrow line ratios.
Supplementary Fig.~4a presents the measured line ratios, in addition to widely used classification schemes\cite{Kewley2001_BPT,Kauffmann03_AGN_HOSTS,Schawinski2007_feedback_ETs} -- that is, the so-called Baldwin, Phillips \& Terlevich (BPT) diagram\cite{BPT1981}.
The emission line ratios we measure, and specifically those measured from the earliest high-resolution spectrum of \mysobj\ taken 10 days after discovery (in the rest-frame), are consistent with being driven by ionizing radiation from a ``composite'' emission source, which includes radiation from an accreting SMBH (that is, an AGN) in addition to young stars (that is, star-forming regions). 
The light-curve of the \oiii/\hbeta\ line ratio, which includes all the available spectra where we could perform our spectral decomposition (that is., among the Las Cumbres Observatory spectra), is shown in Supplementary Fig.~4b.
The light-curve shows essentially no real variability in the line ratio, which is expected given 
the long recombination time-scale in the low-density narrow-line emission region (of order 100 years; ref.~\citen{Peterson2013_N5548_NLR_RM}) 
and the long light travel time-scales ($\gg$100 years; refs.~\cite{Bennert2002_NLR_RL,Mor2009}).

We caution that, with a broad line width of about 2000\,\kms, and given the limited spectral resolution of the Las Cumbres data, it becomes challenging to properly decompose the narrow \hbeta\ line emission from the entire line profile. This is the reason why many measurements are so close to the $\oiii/\hbeta > 1$ constraint that is inherent to our spectral decomposition procedure.

%%%%%%%%%%%%%%%%%%%%%%%%%%%%%%%%%%%%%%%%%%%%%%%%%%%%%%%%%%%%%%%%
\section*{Near-IR spectroscopy.}
\label{SM_sec_obs_spec_nir}
\noindent
We obtained two NIR spectra of \mysobj.
The first spectrum was obtained on 2017 April 21 (i.e., 2 months after the transient detection), with the FIRE instrument on the Magellan-Baade telescope at the Las Campanas observatory\cite{Simcoe2008_FIRE}.
The second spectrum was obtained on May 31 (99 days after detection), with the FLAMINGOS-2 instrument on the Gemini-South telescope\cite{Eikenberry2004_FLAMINGOS2}.
Both spectra are shown in Supplementary Fig.~6, compared to a NIR composite spectrum of 27 quasars, taken from ref.~\citen{Glikman2006_NIR_SED}.
They exhibit a number of broad hydrogen emission lines (for example, ${\rm Pa}\,6\,\lambda 10938.095$, ${\rm Pa}\,7\,\lambda 10049.37$, ${\rm Pa}\,8\,\lambda9545.97$, ${\rm Pa}\,9\,\lambda 9229.01 $,${\rm Pa}\,11\,\lambda 8438.00$), 
as well as strong helium lines, such as $\hei\,\lambda 10829.894$ and $\heii\,\lambda 10123.61$.
Importantly, in \mysobj\ all these emission lines are single-peaked, while the helium features are relatively strong.
This is also the case for other helium emission lines across the optical regime, including He\,{\sc ii} $\lambda3203$ (see Supplementary Fig.~3).
This rules out the possibility that the blue peak within the broad optical emission feature (near 4640 \AA) originates in a disk-like \heii\ emitting region.

%%%%%%%%%%%%%%%%%%%%%%%%%%%%%%%%%%%%%%%%%%%%%%%%%%%%%%%%%%%%%%%%%%%%%%%%%%%%%%%%%%%%%%%%%%%%%%%%%%%%%%%%%%%%%%%%%%%%%%%%%%%%%%%%%%%%%%
\section*{Determination of key SMBH properties}
\label{SM_sec_BH_props}

\noindent
The range of mass accretion rates is derived from bolometric luminosities, that are in turn estimated by assuming the X-ray and optical bolometric corrections of ref.~\citen{Marconi2004}.
For the UV bolometric correction, we assume a value of 3.5, which reflects the range of values reported in the literature\cite{Runnoe2012,Netzer2016_herschel_hiz}, and the expectation that the bolometric correction at $\lambda_{\rm rest}{\simeq}2100$ \AA\ (observed with \swift/UVOT) would be larger than those calibrated for AGN continuum emission at 1450 \AA\ (following a general AGN SED of $f_\nu \propto \nu^{-1/2}$; for example, ref.~\citen{VandenBerk2001}).
We stress that the UV-based bolometric luminosity, of $\approx2.9\times10^{45}\,\ergs$, is probably an overestimate, since the rest of the SED of \mysobj\ does not scale with the UV emission as in normal AGN\cite{Lusso2016_Lx_Luv}.
To obtain mass accretion rates ($\dot{M}$), we finally assume a radiative efficiency of $\eta=0.1$, where $\eta \equiv \Lbol /\dot{M} c^2$ (and $c$ is the velocity of light).

The estimate of the size of the \hbeta-emitting region, $\RBLR(\hbeta)$, and of the BH mass, \mbh, relies on the results of reverberation mapping experiments of broad-line AGN, which provide a way to link the observed optical continuum luminosity and the size of the line-emitting region and to obtain single-epoch, ``virial'' mass estimates\cite{Kaspi2000,Shen2013_rev,Peterson2014_review,Mejia2016_XS_MBH}.
For \mysobj, the working assumption of a virialized BLR (or \hbeta-emitting region) can be justified by the (persistent) shapes of the broad Balmer emission lines, which resemble those of normal broad-line AGN.
Moreover, the optical luminosities measured from our spectroscopy place \mysobj\ at the heart of the range of luminosities probed by reverberation mapping experiments of low-redshift AGN, and of the corresponding BLR size -- luminosity relations\cite{Bentz2013_lowL_RL}.
In what follows, we rely on the best-fit spectral model for the \hbfitdate, \hbfittel/\hbfitinst\ spectrum (described in 'Optical spectroscopy' above).
The monochromatic luminosity at 5100 \AA, of $\lambda L_{\lambda}(5,100\,{\rm \AA}){=}\Loptfourthree{\times}10^{43}\,\ergs$, translates to an \hbeta-emitting region size of $\RBLR=21.9$ or $28.4$ light-days, following the prescriptions given in refs.~\citen{TrakhtNetzer2012_Mg2} and \citen{Bentz2013_lowL_RL}, respectively.
Combining the former value with an \hbeta\ line width of $\fwhb{=}\hbfitfw\,\kms$, and the mass prescription given in ref.~\citen{TrakhtNetzer2012_Mg2}, we obtain $\mbh{=}\Mseven{\times}10^{7}\,\Msol$. 
The larger (latter) \RBLR\ estimate would increase this \mbh\ estimate by about 0.1 dex, which is much smaller than the systematic uncertainties on \mbh, which are of order 0.3-0.4 dex (ref.~\citen{Shen2013_rev}).
Given the extremely intense UV continuum emission and the fact that the size of the \hbeta-emitting region is driven by the ionizing UV (rather than optical) continuum, we further caution that the BLR size, and thus \mbh, could be considerably larger, perhaps by as much as a factor of ${\sim}4$ 
(see, for example, refs.~\citen{Kaspi2005,Pei2017_N5548_RM}).

% The same spectral decomposition procedure was also used to measure the widths and fluxes of the two features comprising the double-peaked feature near 4680 \AA. 
% Each of the peaks was modelled with a single Gaussian profile, with a width similar to (the core of) the broad \hbeta\ emission line. 
% We stress that the spectral decomposition procedure accounts for the much weaker, blended emission from ionized iron (see ref.~\citen{TrakhtNetzer2012_Mg2} for details).

%%%%%%%%%%%%%%%%%%%%%%%%%%%%%%%%%%%%%%%%%%%%%%%%%%%%%%%%%%%%%%%%%%%%%%%%%%%%%%%%%%%%%%%%%%%%%%%%%%%%%%%%%%%%%%%%%%%%%%%%%%%%%%%%%%%%%%
% \section{Bowen Fluorescence in Active Galactic Nuclei}
% \label{SM_sec_BF}

%%%%%%%%%%%%%%%%%%%%%%%%%%%%%%%%%%%%%%%%%%%%%%%%%%%%%%%%%%%%%%%%%%%%%%%%%%%%%%%%%%%%%%%%%%%%%%%%%%%%%%%%%%%%%%%%%%%%%%%%%%%%%%%%%%%%%%
\section*{Relevant mechanisms for the long-lived UV flare}
\label{SM_sec_mechanisms}

\noindent
Here We discuss some of the mechanisms that may be considered relevant for driving the intense UV brightening in \mysobj\ and the other events we associate with this proposed class.
As noted in the main text, given the evidence for AGN-like activity in \mysobj\ (and the other events) both before and after the UV-optical event, we focus our brief discussion here on processes related to (thin) accretion disks that feed SMBHs.
In this context, we note that the transient optical rise time of a few weeks is only slightly longer than the dynamical timescale in a thin disk, given our \mbh\ estimate ($\sim$4 days; see \S\ref{SM_sec_timescales} below).
The longevity of the enhanced UV emission, of over a year, is considerably longer than the thermal timescale ($\sim$50 days, at most), and is instead starting to be comparable with the time-scales over which heating fronts may propagate through the disk ($\sim$3 years).

\noindent
{\bf Tidal disruption events.}
Some TDEs, which are generally thought to be powered by a newly formed, transient accretion disk, show a fast increase in UV continuum emission, accompanied by strong \HeIIop\ emission\cite{Gezari2012_TDE,Arcavi2014_TDEs_He,Holoien2016_TDE_AS15oi,Holoien2016_TDE_AS14li,Blagorodnova2017_TDE_iPTF16fnl,Hung2017_TDE_iPTF16axa}.
However, the temporal evolution of \mysobj, F01004-2237 and OGLE17aaj is much slower than what is seen in TDEs (see Fig.~\ref{fig:img_spec_monitoring} and Supplementary Fig.~1).
In addition, the \HeIIop\ line seen in these three transients is much narrower than in TDEs (see Fig.~\ref{fig:heii_spec_comp_tdes}).
Although the TDE ASASSN-14li displays emission lines that are narrower than other TDEs (in a late-time spectrum, 86-days from discovery; ref.~\citen{Holoien2016_TDE_AS14li}), it still exhibits a very broad base of the \heii\ line, not seen in our class of events. 
Tidal disruptions of giant stars can produce slowly-evolving light curves
\cite{MacLeod2012_TDEs_giants}, but in that case both the rise and decline of the light curve are expected to last several years, whereas the rise in our class of events is much more sudden.
Recently, two events claimed to be TDEs were reported to have long-lived light curves. 
One is interpreted to be a TDE around an intermediate-mass BH\cite{Lin2018_TDE_IMBH} and the other is a dust-reprocessed TDE ``echo'' in a merging starburst galaxy\cite{Mattila2018_TDE_Arp299}. Neither of these cases relate to \mysobj\ (nor to the two sources we associate with it).
Moreover, the strong \OIII\ line emission seen in all three events suggests that the SMBHs were not completely dormant prior to the sudden (UV/optical) brightening, 
% unlike what is considered for TDEs. 
further weakening the case for a TDE.
Given these stark differences, we conclude that a TDE is unlikely to be driving \mysobj\ (and by association, the F01004-2237 event and OGLE17aaj).

\noindent
{\bf An interaction between an outflow and the BLR.}
A recently published model describes the interaction of an outflow launched from the vicinity of a SMBH with the BLR gas\cite{Moriya2017_BLR_winds_transients}. 
According to this model, an outflow with $v \lesssim 0.3\,c$ may account for the $>$1 year-long enhanced emission we observe in \mysobj. 
Such an outflow would, however, require a period of several months ($\RBLR / 0.1\,c$) to reach the BLR, and it is unclear how this mechanism would provide favourable conditions for extreme UV and BF emission.
If the broad \hbeta\ line is also emitted from an out-flowing region, then this  may have implications for our estimates of \mbh\ and \lledd, as they assumed a virialized \hbeta\ line emitting region (see ref.~\citen{Netzer_Marziani2010} for a detailed discussion).

\noindent
{\bf Other mechanisms.}
There are several other physical mechanisms that have been proposed to give rise to dramatic increase in (UV-optical) emission from accreting SMBHs.
Periodic tidal interactions between the SMBH and an orbiting binary pair of stars may account for recurring ``outbursts'' of accretion-driven X-ray and optical emission, separated by several years, like those observed in some AGN\cite{Campana2015_IC3599,Grupe2015_IC3599}.
In each interaction episode some of the gas in the star envelope is accreted onto the SMBH through a transient disk, 
perhaps masquerading as a TDE\cite{MetzgerStone2017_TDE_impostors}.
A stellar origin for the gas may explain the high metallicity and density of the BF line-emitting gas in \mysobj. 
However, it is unclear how this scenario would explain the persistence of the current episode of enhanced accretion (and UV emission) lasting over a year or, more importantly, the emission line profiles, which indicate an extended distribution of highly-ionized gas.
Some models suggest that binary SMBHs may also exhibit sudden enhancement of SMBH accretion, accompanied by increased UV emission, driven by occasional shocks and events of drastic angular momentum loss in the dense circum-binary accretion flow\cite{Farris2015_BBHs_Acc}.
As these episodes are expected to last over periods that are several times the binary orbital period, the prolonged enhanced UV emission can be accounted for, unless the binary separation is extremely tight (that is $a\lesssim 0.005$ pc).
Given the data in hand, however, it is impossible to determine the real state of accretion onto the SMBH in \mysobj\ over the periods required to test any of these scenarios, and particularly between the archival detections (in the X-rays and UV) and the recent increase in UV-optical flux.

%%%%%%%%%%%%%%%%%%%%%%%%%%%%%%%%%%%%%%%%%%%%%%%%%%%%%%%%%%%%%%%%%%%%%%%%%%%%%%%%%%%%%%%%%%%%%%%%%%%%%%%%%%%%%%%%%%%%%%%%%%%%%%%%%%%%%%
\smallskip
\section*{Typical timescales in thin accretion disks around SMBHs}
\label{SM_sec_timescales}

% \smallskip
\noindent
For the sake of completeness, we recall here several key timescales in geometrically-thin, optically-thick accretion disks around AGN (and SMBHs).
These timescales are derived in many references (for example, refs.~\citen{Frank2002_book_ACC,Netzer2013_book}).
% {\color{asparagus}CR: I would not use "countless" references.}
Here we follow the recent discussion of these timescales in the context of changing look AGN, provided in ref.~\citen{Stern2018_CLAGN_WISE}.
The timescales are commonly parametrized in terms of 
the BH mass, \mbh\ (which for \mysobj\ we estimate to be ${\sim}1.4{\times}10^7\,\Msol$); 
the distance from the BH, in terms of the gravitational radius, $r/r_{\rm g}$ (where $r_{\rm g} \equiv G\mbh/c^2$); 
the disk scaled height $h/r$, which in thin disks is of order $\sim$0.05; 
and the pseudo-viscosity parameter $\alpha$, which in thin disks is of order $\sim$0.03.
The dynamical timescale is the typical timescale for (azimuthal) Keplerian motion of the disk material around the BH, and is given by:
\begin{align}
t_{\rm dyn} &\simeq 14\,{\rm hours}  
\left(\frac{\mbh}{10^7\,\Msol}\right) 
\left(\frac{r}{100\,r_{\rm g}}\right)^{3/2}
\label{SM_eq_t_dyn}
\end{align}
\noindent
The thermal timescale is the typical timescale for the disk cooling or heating, and thus further depends on the viscosity parameter:
\begin{align}
t_{\rm th} &= t_{\rm dyn}/\alpha 
{\simeq} 19\,{\rm days}
\left(\frac{\mbh}{10^7\,\Msol}\right)
\left(\frac{r}{100\,r_{\rm g}}\right)^{3/2}
\left(\frac{\alpha}{0.03}\right)^{-1} .
\label{SM_eq_t_th}
\end{align}
\noindent
Cooling or heating fronts may travel throughout the disk on longer timescales, accounting for the disk geometry:
\begin{align}
t_{\rm front} &= t_{\rm th}/(h/r) 
{\simeq} 380\,{\rm days} \nonumber \\
& \quad \left(\frac{\mbh}{10^7\,\Msol}\right)
\left(\frac{r}{100\,r_{\rm g}}\right)^{3/2}
\left(\frac{\alpha}{0.03}\right)^{-1}
\left(\frac{h/r}{0.05}\right)^{-1} .
\label{SM_eq_t_front}
\end{align}
\noindent
Finally, the viscous timescale, over which material travels radially from a radius $r$ to the BH, is yet longer:
\begin{align}
t_{\rm \nu} &= t_{\rm front}/(h/r) 
{\simeq} 21\,{\rm years} \nonumber \\
& \quad \left(\frac{\mbh}{10^7\,\Msol}\right) 
\left(\frac{r}{100\,r_{\rm g}}\right)^{3/2} 
\left(\frac{\alpha}{0.03}\right)^{-1} 
\left(\frac{h/r}{0.05}\right)^{-2} .
\label{SM_eq_t_visc}
\end{align}
\noindent
We recall that the radius within the disk that is relevant for the continuum emission is both wavelength- and accretion rate- dependent, so that the dynamical timescale can be expressed as: 
$t_{\rm dyn} {\simeq} 0.5\,{\rm sec}\, \dot{M}^{1/2} \lambda^{2}$ 
(with $\dot{M}$ given in \mpyr\ and $\lambda$ in \AA).
Given the derived properties of \mysobj, the near-UV data (with $\lambda_{\rm Eff} \simeq 2200$ \AA) would correspond to ${\sim}300\,r_{\rm g}$.

%%%%%%%%%%%%%%%%%%%%%%%%%%%%%%%%%%%%%%%%%%%%%%%%%%%%%%%%%%%%%%%%%%
%%%%%%%%%%%%%%%%%%%%%%%%%%%%%%%%%%%%%%%%%%%%%%%%%%%%%%%%%%%%%%%%%%

%%%%%%%%%%%%%%%%%%%%%%%%%%%%%%%%%%%%%%%%%%%%%%%%%%%%%%%%%%%%%%%%%%
% DATA AVAILABILITY
%%%%%%%%%%%%%%%%%%%%%%%%%%%%%%%%%%%%%%%%%%%%%%%%%%%%%%%%%%%%%%%%%%

\smallskip
\paragraph{Data availability}
The data that support the findings of this study are available from the corresponding author upon reasonable request.
All of our spectra are publicly available on the Weizmann Interactive Supernova data REPository (WISeREP)\cite{YaronGalYam_2012_WISeREP}.
The data used to prepare Supplementary Fig.~1 are available from the ASAS-SN Light Curve Server (\url{https://asas-sn.osu.edu/}).

%%%%%%%%%%%%%%%%%%%%%%%%%%%%%%%%%%%%%%%%%%%%%%%%%%%%%%%%%%%%%%%%%%
% METHODS REFS
%%%%%%%%%%%%%%%%%%%%%%%%%%%%%%%%%%%%%%%%%%%%%%%%%%%%%%%%%%%%%%%%%%

\paragraph{Additional References}

\clearpage
\newpage

%%%%%%%%%%%%%%%%%%%%%%%%%%%%%%%%%%%%%%%%%%%%%%%%%%%%%%%%%%%%%%%%%%%%%%

\clearpage
\pagestyle{empty}
\setboolean{@twoside}{false}


\section*{Supplementary material}

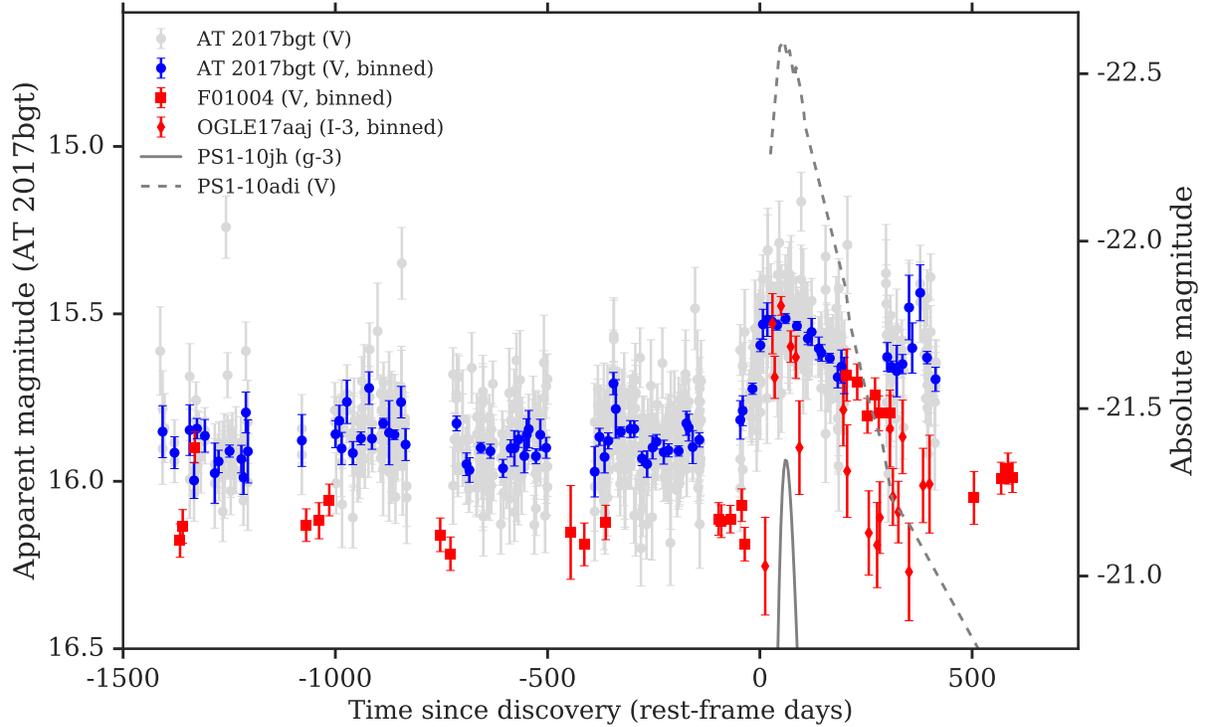

Supplementary Figure 1: **The transient nature and the long-term optical light-curve of AT 2017bgt and comparison objects.** We show both the raw ASAS-SN $V$-band data for AT 2017bgt (grey symbols), and a binned light-curve (blue symbols; binned in 4-day time spans, and cleaned from data points with errors $>0.1$ mag). We also show the binned optical $V$-band light-curves of the other two transients we associate with AT 2017bgt – the event reported in F01004-2237 (ref. 1), and the event OGLE17aaj (refs. 2, 3). In all three cases, the magnitudes are calibrated in the Vega system, and the error bars on the binned light-curves represent a combination of uncertainties on individual measurements and the scatter of data in each bin. For comparison, we present light-curves of two other recently reported events (PS1-10jh, solid line, ref. 4 and PS1-10adi, dashed line, ref. 5), which are markedly different than AT 2017bgt and the two source we associate with it. All magnitudes are in the AB system.

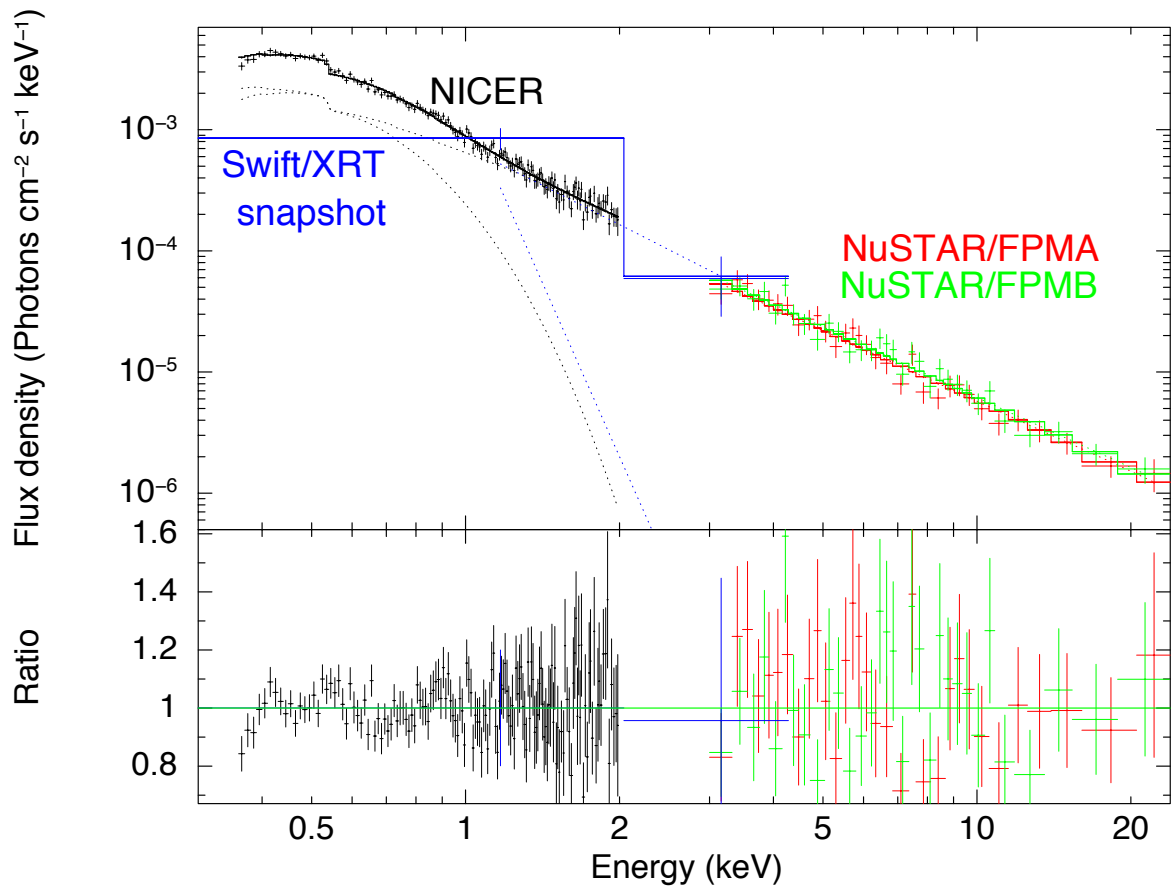

Supplementary Figure 2: **The X-ray spectral energy distribution of AT 2017bgt.** The blue, black, red and green data points (and uncertainties) correspond to the *NICER*, *Swift*/XRT, and *NuSTAR* FPMA/FPMB data, respectively, taken quasi-simultaneously starting 2018 June 25. All error bars indicate $1-\sigma$ uncertainties. The solid line traces the best-fit model, which consists of an absorbed power-law and a blackbody component. The bottom panel shows the residuals of the fit.

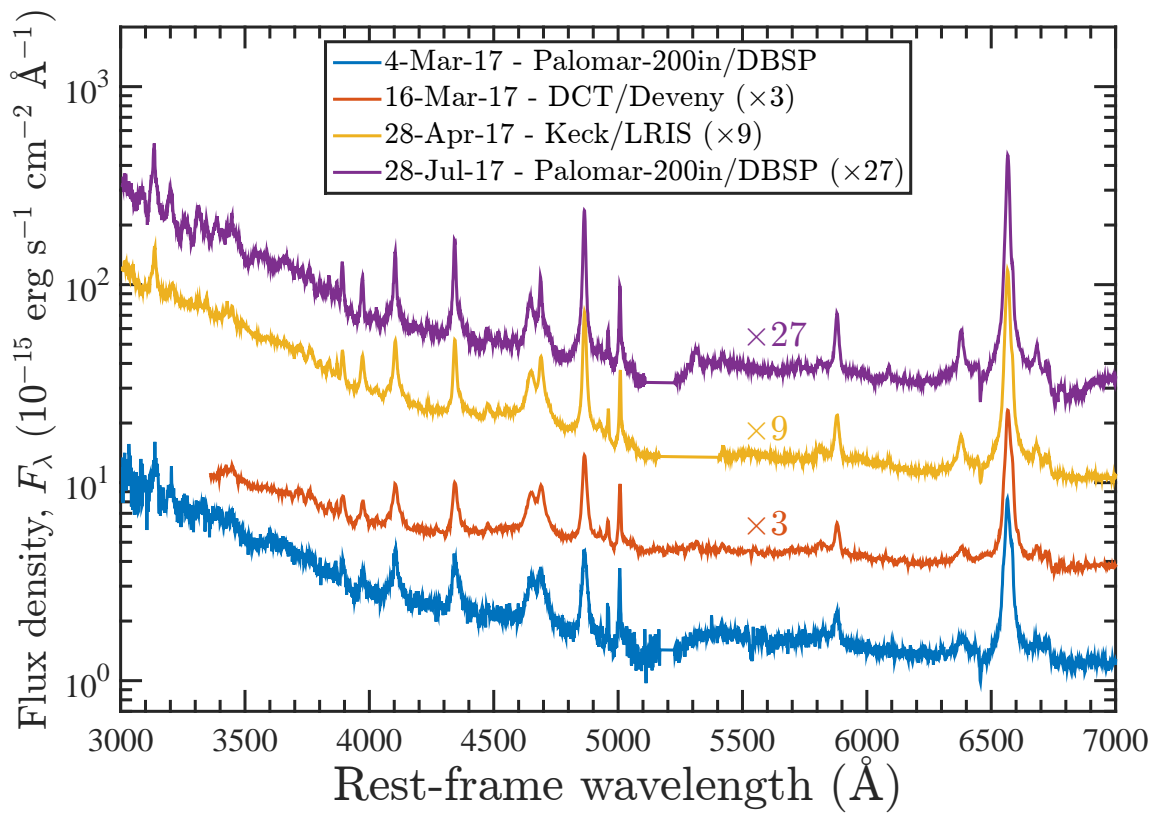

Supplementary Figure 3: **High-resolution optical spectra of AT 2017bgt.** We show four spectra taken on 2017 March 3, March 16, April 28, and July 28 (10, 23, 66, and 157 days after discovery; blue, orange, yellow, and purple lines, respecrively). The spectra were obtained using the DoubleSpec instrument on the Hale 200 inch telescope at the Palomar observatory (first and last spectra); the DeVeny instrument on the Discovery Channel Telescope at the Lowell Observatory; and the LRIS instrument on the W. M. Keck telescope at Mauna Kea. All spectra are presented un-binned and scaled by a multiplicative factor (consecutive steps of ×3, as indicated). All spectra clearly show the emission features discussed in throughout this study, including broad Balmer lines, forbidden narrow lines, and the double-peaked broad feature near 4640 Å.

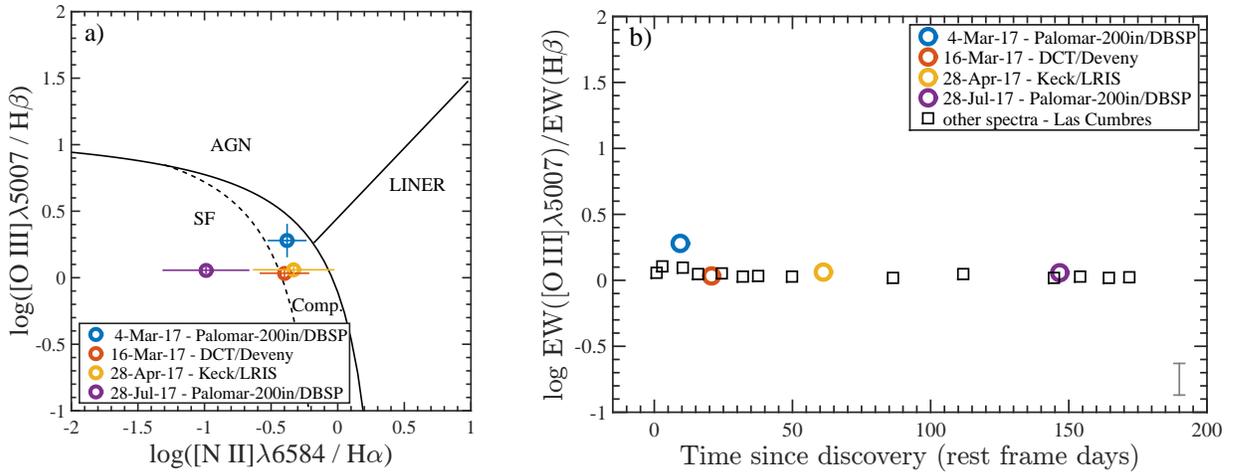

Supplementary Figure 4: **Strong emission line ratio diagnostics for AT 2017bgt.** In panel **a** we show the [O III] $\lambda$5007/H$\beta$ vs. [N II] $\lambda$6584/H$\alpha$ line ratios, as measured from the higher spectral resolution, high-S/N spectra presented in Supplementary Figure 3. The different black lines, taken from refs. 6–8, illustrate commonly used boundaries between sources classified as star forming regions/galaxies (SF), AGN, low ionization nuclear emission line regions (LINERs), and "composite" sources. Panel **b** shows the light-curve of the [O III] /H$\beta$ line ratio, measured from all the available optical spectra for which this measurement was possible, with data from the four high-resolution spectra highlighted. The grey error bars at the bottom-right illustrates the typical (median) uncertainty on the line-ratio measurements (about 0.1 dex). Our earliest post-discovery optical spectra of AT 2017bgt suggest that the narrow emission lines are photoionized by a composite of continuum sources, which include both SF- and AGN-like emission.

5

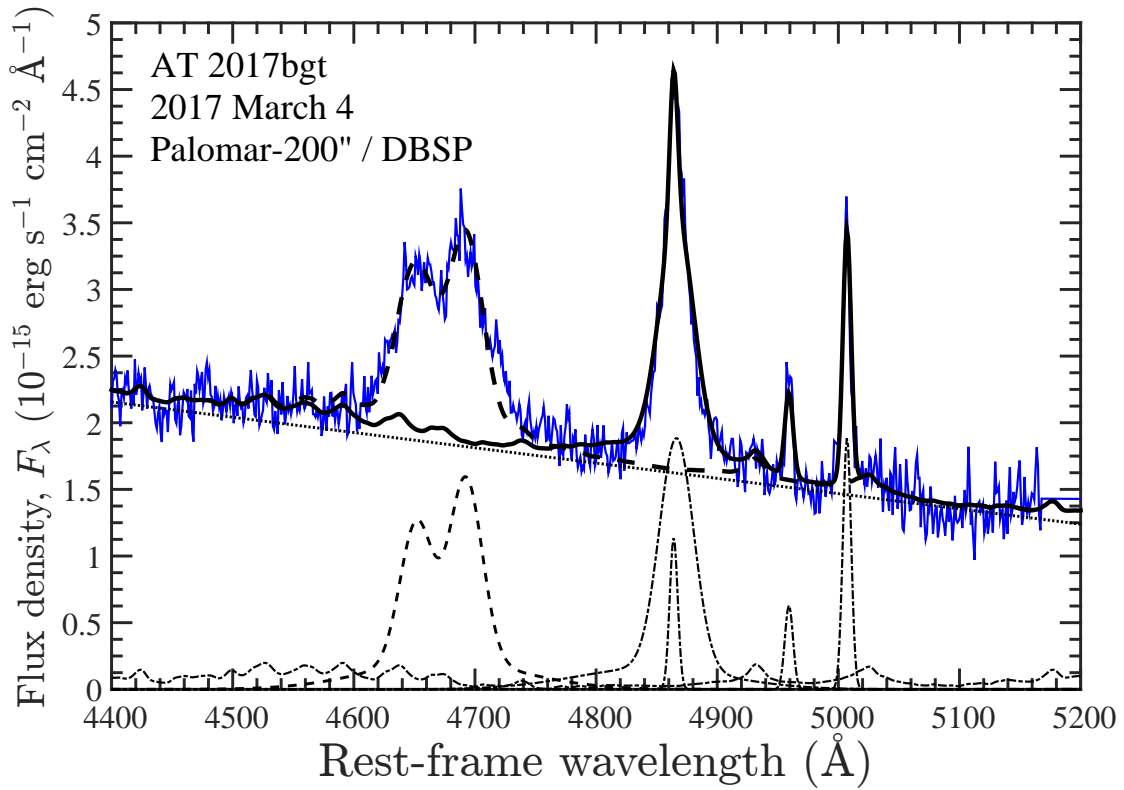

Supplementary Figure 5: **Spectral decomposition of the optical spectrum of AT 2017bgt.** The 2017 March 4 optical spectrum of AT 2017bgt, taken with the DoubleSpec instrument mounted on the Hale 200-inch telescope at the Palomar observatory, is shown in blue. The dotted diagonal line traces the best-fit (linear) continuum emission. The lines at the bottom trace the different components of the spectral model. Dot-dashed lines trace the broad H$\beta$ emission line, the narrow [O III] emission lines, and the broadened template of iron emission features. The combination of these components, typically seen in unobscured AGN, is shown in a solid heavy black line. The dashed lines highlight the two additional emission line profiles, which are used to decompose the broad, double-peaked emission feature near $\sim$4640 Å.

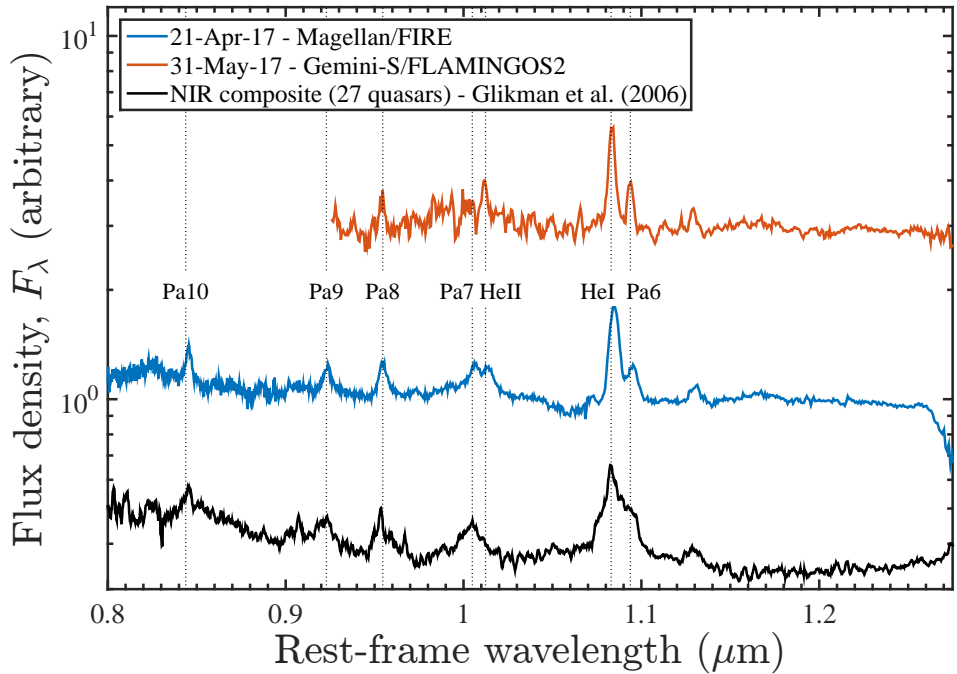

Supplementary Figure 6: **Near-infrared spectroscopy of AT 2017bgt.** The two spectra were obtained on 2017 April 21 and May 31 (59 and 99 days after the transient detection; blue and red lines, respectively), using the FIRE instrument on the Magellan-Baade telescope at the Las Campanas observatory, and the FLAMINGOS-2 instrument on the Gemini-South telescope. Here we show the $J$-band portion of the spectra, with annotated emission lines from hydrogen (Paschen series), He I, and He II, compared to a composite NIR quasar spectrum taken from ref. 9. All spectra are normalized at 1.15 $\mu$m, and then shifted by a multiplicative factor (consecutive steps of $\times 3$). Compared to normal unobscured AGN, AT 2017bgt exhibits strong helium emission lines, but no signature of double-peaked hydrogen emission features.

Supplementary Table 1: **Key measured and derived properties of AT 2017bgt**. The table lists the source of the data used, the date of the observation, and the value associated with the measurement.

|  | property | data source | date | value |
| --- | --- | --- | --- | --- |
| Archival | $\nu L_\nu$(NUV) | *GALEX* | 2004 May 17 | $1.2 \times 10^{43}\,\mathrm{erg\,s^{-1}}$ |
|  | $L(2-10\,\mathrm{keV})$ | *ROSAT* | 1990 Aug. 9 | $5.3 \times 10^{42}\,\mathrm{erg\,s^{-1}}$ |
| New, post-transient | $\nu L_\nu$(NUV) | *Swift*/UVOT | 2017 April 24 | $8.9 \times 10^{44}\,\mathrm{erg\,s^{-1}}$ |
|  | $L(2-10\,\mathrm{keV})$ | *Swift*/XRT | 2017 April 24 | $1.2 \times 10^{43}\,\mathrm{erg\,s^{-1}}$ |
|  | $\nu L_\nu$(5100 Å) | Palomar spec. | 2017 March 4 | $7.3 \times 10^{43}\,\mathrm{erg\,s^{-1}}$ |
| Derived | $L_{\mathrm{bol}}$ (from 2-10 keV) | *Swift*/XRT | 2017 April 24 | $2.2 \times 10^{44}\,\mathrm{erg\,s^{-1}}$ |
|  | $L_{\mathrm{bol}}$ (from 5100 Å) | Keck spec. | 2017 March 4 | $5.8 \times 10^{44}\,\mathrm{erg\,s^{-1}}$ |
|  | $M_{\mathrm{BH}}$ | Palomar spec. | 2017 March 4 | $1.8 \times 10^{7}\,M_\odot$ |
|  | $L/L_{\mathrm{Edd}}$ (from 2-10 keV) | Palomar & XRT | $\cdots$ | 0.08 |
|  | $L/L_{\mathrm{Edd}}$ (from 5100 Å) | Palomar spec. | 2017 March 4 | 0.21 |

**References cited in this Supplementary Information file**



\end{document}